\documentclass[aps,prl,reprint,superscriptaddress]{revtex4-1}
\usepackage{amsmath}
\usepackage{amssymb}
\usepackage{amsfonts}
\usepackage{bbm}
\usepackage{bm}
\usepackage{graphicx}
\usepackage{color}
\usepackage{dcolumn}
\usepackage{hyperref}

\begin{document}

\title{Universal topological quench dynamics: Altland-Zirnbauer tenfold classes}

\author{Lin Zhang}
\affiliation{International Center for Quantum Materials and School of Physics, Peking University, Beijing 100871, China}
\affiliation{Collaborative Innovation Center of Quantum Matter, Beijing 100871, China}
\author{Wei Jia}
\affiliation{International Center for Quantum Materials and School of Physics, Peking University, Beijing 100871, China}
\affiliation{Collaborative Innovation Center of Quantum Matter, Beijing 100871, China}
\author{Xiong-Jun Liu}
\thanks{Corresponding author: xiongjunliu@pku.edu.cn}
\affiliation{International Center for Quantum Materials and School of Physics, Peking University, Beijing 100871, China}
\affiliation{Collaborative Innovation Center of Quantum Matter, Beijing 100871, China}
\affiliation{CAS Center for Excellence in Topological Quantum Computation, University of Chinese Academy of Sciences, Beijing 100190, China}
\affiliation{Institute for Quantum Science and Engineering and Department of Physics, Southern University of Science and Technology, Shenzhen 518055, China}

\date{\today}

\begin{abstract}
Topological phases of the famous Altland-Zirnbauer (AZ) tenfold classes are defined on the equilibrium ground states. Whether such equilibrium topological phases have universal correspondence to far-from-equilibrium quantum dynamics is a fundamental issue of both theoretical and experimental importance. Here we uncover the universal topological quench dynamics linking to the equilibrium topological phases for the complete AZ tenfold classes, with a general framework being established. We show a fundamental result that a $d$-dimensional topological phase of the tenfold class, with an integer invariant or $\mathbb{Z}_{2}$ index defined on high symmetry momenta, is generically characterized by topology reduced to the highest-order band-inversion surfaces located at arbitrary discrete momenta of Brillouin zone. Such dimension-reduced topology is further captured by universal topological patterns emerging in far-from-equilibrium quantum dynamics by quenching the system from trivial phase to the topological regime, 
rendering the dynamical hallmark of the equilibrium topological phase. 
This work establishes a universal dynamical characterization for the complete AZ symmetry classes of topological phases, which has broad applications in theory and experiment.
\end{abstract}

\maketitle

\emph{\bf Introduction}.---Topological quantum phases of the famous Altland-Zirnbauer (AZ) symmetry classes are defined on the ground states of free-fermion quantum systems at equilibrium~\cite{Hasan2010,Qi2011,Chiu2016}, and are classified by the tenfold way based on the space dimension and the non-spatial time-reversal (TR), particle-hole (PH), and/or chiral symmetries~\cite{Chiu2016,Schnyder2008,Kitaev2009,Ryu2010}. Those fundamental topological phases can be characterized by the equilibrium properties such as the bulk topological invariants, the protected gapless boundary modes~\cite{Chiu2016}, the Wilson loop~\cite{Yu2011}, and the entanglement spectra~\cite{Fidkowski2010,Turner2010}.

The notion of topological phases is not restricted in equilibrium systems. In the periodic driving and quantum quenches, the anomalous Floquet topological phases~\cite{Rudner2013,Harper2020,Rudner2020,Wintersperger2020} and dynamical topological phases or phase transitions~\cite{Vajna2015,Heyl2018,Flaschner2018,Gong2018,McGinley2019} are uncovered, revealing the exotic far-from-equilibrium responses of topological states.
Particularly, a universal correspondence was predicted between equilibrium topological phases of integer invariants and nonequilibrium dynamical topology induced in quenching the systems across topological transitions
~\cite{Zhang2018,Zhang2020,Yu2020}. This prediction on one hand opens a generic way to characterize equilibrium topological phases by far-from-equilibrium quantum dynamics, facilitating the detection of topological states by quantum dynamics~\cite{Wang2017,Tarnowski2019,Mizoguchi2021}, and on the other hand, shows new insight into classifying nonequilibrium quantum dynamics by topological theory. 
The follow-up experimental studies~\cite{Sun2018,Yi2019,Song2019,Ji2020,Xin2020,Niu2020,BChen2021} and further theoretical generalizations~\cite{Zhou2018,Zhu2020,Hu2020,Li2020,Ye2020,Yu2021} have been widely reported.
Nevertheless, thus far all the developed theory is only applicable to integer-valued topological phases, while whether a general framework for the complete AZ classes of topological phases can be established is an open question. 
The major impediment is that, the full AZ classes cover the $\mathbb{Z}_{2}$ topological phases, e.g. the TR invariant topological insulators, whose invariants are usually defined on highly symmetric momenta~\cite{Fu2006,Fu2007}, while the dynamical characterization theory necessitates to reduce the bulk topology to band inversion surfaces (BISs), a key concept of the theory and located on arbitrary momenta in the Brillouin zone (BZ)~\cite{Zhang2018,Yu2020}. Whether the $\mathbb{Z}_{2}$ phases can be characterized via BISs was an unresolved outstanding issue.

Here we establish the first dynamical characterization for the complete tenfold symmetry classes of topological phases by uncovering universal topological quench dynamics linking to the AZ equilibrium topological phases. 
The key finding is that an AZ class $d$-dimensional ($d$D) topological phase with an integer or $\mathbb{Z}_{2}$ invariant can be characterized by a $0$D invariant defined on the highest order BISs located at arbitrary discrete momenta. Further, we show that the dimension-reduced topology on BISs can be further captured by the universal topological patterns emerging in the quench dynamics. The broad applications of theory are commented and illustrated.

\emph{\bf Generic model of AZ symmetry classes}.---We describe the generic $d$D topological phase of AZ symmetry classes with the elementary Dirac Hamiltonian
\begin{equation}\label{eq:Hamiltonian}
    H(\mathbf{k}) = \mathbf{h}(\mathbf{k})\cdot\boldsymbol{\gamma} = \sum_{i=0}^{d}h_{i}(\mathbf{k})\gamma_{i}+\sum_{i=d+1}^{d'}h_{i}(\mathbf{k})\gamma_{i},
\end{equation}
where $\mathbf{k}$ is the $d$D momentum, with $d'=d$ for integer topological phases and $d'=d+1$ ($d+2$) for the first (second) descendant $\mathbb{Z}_{2}$ topological phases (Tab.~\ref{tab:table1})~\cite{Ryu2010,Qi2008}. The Clifford algebra $\gamma$ satisfies anti-commutation relation $\{\gamma_{i},\gamma_{j}\}=2\delta_{ij}$, mimicking a (pseudo)spin of dimensionality $n_{d'}=2^{d'/2}$ ($2^{(d'+1)/2}$) if $d'$ is even (odd).
For instance, the $2$D, $3$D and $4$D TR-invariant topological insulators in class AII involve at least four bands and can be represented by five Dirac matrices.

The complex AZ classes of topological phases are all classified by integer invariants. For the real AZ classes, the coefficients $h_{i}(\mathbf{k})$ of Dirac Hamiltonian \eqref{eq:Hamiltonian} are constrained by the corresponding TR and/or PH symmetries, and have to be either odd or even with respect to $\mathbf{k}$, as detailed in Supplemental Material~\cite{SM}.

\begin{table}[t]
    \caption{\label{tab:table1}%
    {\bf Universal dynamical correspondence for the AZ symmetry classes.} Here $d$ is the space dimension, and $s=0,1,\dots,7$ represent the real AZ classes AI, BDI, D, DIII, AII, CII, C, and CI, respectively. The topological phases in equilibrium classification (second row) can be characterized by a dynamical integer invariant $\mathcal{W}$ or $\mathbb{Z}_{2}$ invariant $\nu^{(1,2)}$ (third row). The complex AZ classes A and AIII (not shown here) are characterized by $\mathcal{W}$ dynamically.
    }
    \begin{ruledtabular}
        \begin{tabular}{cccccccccc}
            $d-s\mod 8$ & 1 & 2 & 3 & 4 & 5 & 6 & 7 & 8 \\
            \colrule
            Classification & 0 & 0 & 0 & $2\mathbb{Z}$ & 0 & $\mathbb{Z}^{(2)}_{2}$ & $\mathbb{Z}^{(1)}_{2}$ & $\mathbb{Z}$ \\
            \begin{tabular}{@{}c@{}}Dynamical \\ characterization\end{tabular} & 0 & 0 & 0 & $\mathcal{W}$ & 0 & $\nu^{(2)}$ & $\nu^{(1)}$ & $\mathcal{W}$ \\
        \end{tabular}
    \end{ruledtabular}
\end{table}

\begin{figure}[t]
    \includegraphics{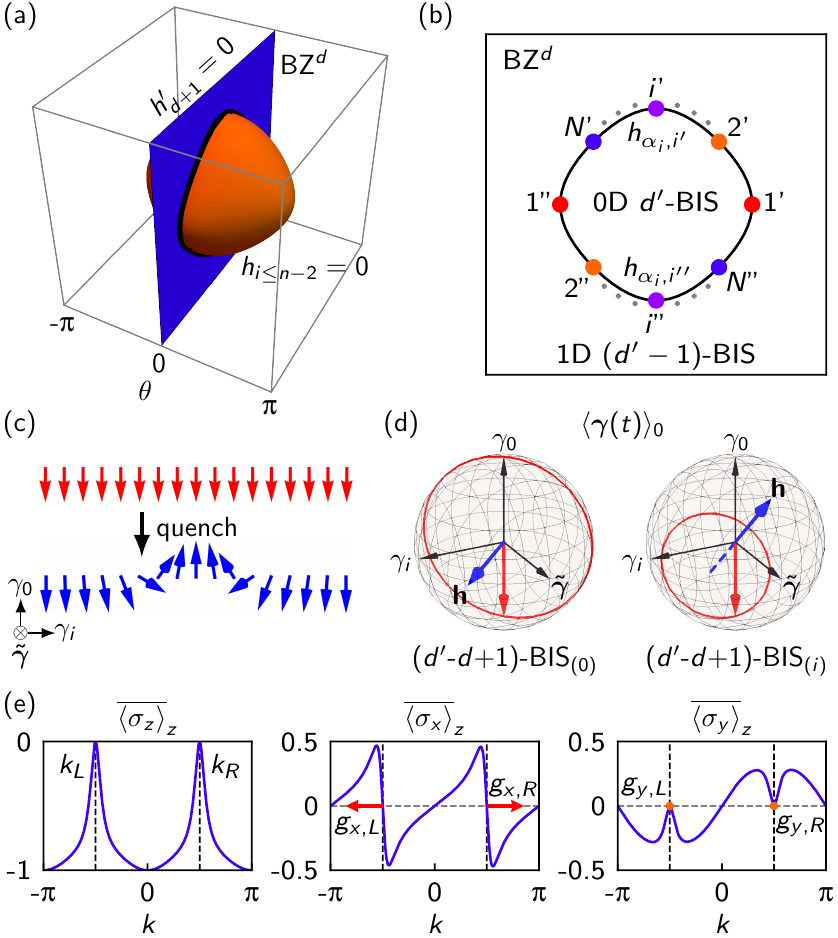}
    \caption{\label{fig:figure1}
    {\bf Dynamical characterization of $\mathbb{Z}_{2}$ topological phases.} (a) Higher-order BISs. The intersection of the hyperplane with $h'_{d+1}=0$ (blue) and the hypersurface with $h_{i}=0$ for $0\leq i\leq n-2$ (orange) gives the $n$-th order BIS for the first descendant $\mathbb{Z}_{2}$ classifications. (b) $\mathbb{Z}_{2}$ topological index on the highest order BISs. The $0$D $d'$-BIS points (colors) on the $1$D $(d'-1)$-BIS (black line) and with $h_{d-1}=0$ are grouped into $N$ symmetric point pairs $(i',i'')$, to each of which a nonzero $h_{\alpha_{i}}$ is assigned. (c) A trivial state fully polarized in the negative direction of axis $\gamma_{0}$ is suddenly quenched to a topological phase. (d) Spin dynamics on the BISs. On the $(d'-d+1)\text{-BIS}_{(0)}$, the spin polarization $\langle\boldsymbol{\gamma}(t)\rangle_{0}$ (red) precesses within the plane perpendicular to the post-quench Hamiltonian $\mathbf{h}$ (blue arrow), leading to the vanishing time averages $\overline{\langle\boldsymbol{\gamma}\rangle}_{0}$. On the $(d'-d+1)\text{-BIS}_{(i\neq 0)}$, $\overline{\langle\boldsymbol{\gamma}\rangle}_{0}$ vanishes only in the $\gamma_{i}$ direction, while all other components are nonzero in general. (e) Numerical results for quenching the $1$D topological phase of class D from $\mu_{\mathrm{i}}=25t_{0}$ to $\mu_{\mathrm{f}}=0$. The vanishing polarization $\overline{\langle\boldsymbol{\sigma}\rangle}_{z}$ gives the BIS points $k_{L,R}$ (black dashed lines), on which the opposite dynamical field $g_{x}$ (red arrows) manifests the nontrivial dynamical $\mathbb{Z}_{2}$ invariant $\nu^{(1)}=-1$, while $g_{y}$ (orange dots) vanishes. Here we set $\Delta=0.2t_{0}$.
    }
\end{figure}

\emph{\bf Topological index on high-order BISs}.---
We show the key result that for AZ ten-fold classes the topology of any topological phase with $\mathbb{Z}_{2}$ or integer invariant can be characterized by the $0$D invariant defined on the highest order BISs. The concept of high-order BISs was introduced with a dimension reduction approach~\cite{Zhang2018,Yu2020}. The first-order BIS, as denoted by $1$-BIS, are a $(d-1)$D momentum subspace, on which one of the components in $H$, say $h_0(\bold k)$, vanishes. Then the $n$-BIS are defined on the $(n-1)$-BIS by setting $h_{n-1}(\bold k)=0$, given by $n\text{-BIS}=\{\mathbf{k}\in\mathrm{BZ}\vert h_{i}(\mathbf{k})=0,i=0,1,\dots,n-1\}$.
For a $d$D topological phase with integer invariant, the topology can be reduced to arbitrary $n$-th order BISs and particularly,
on the highest order $d\text{-BIS}$ which consists of several pairs of points, the integer invariant reads $\mathcal{W}=\sum_{d\text{-BIS}_{j}}[\mathrm{sgn}(h_{d,R_{j}})-\mathrm{sgn}(h_{d,L_{j}})]/2$, with $L_{j}$ ($R_{j}$) being the left (right) hand point of the $j$-th pair on $d\text{-BIS}_{j}$. This is a natural high-order generalization of the BIS-bulk duality for integer topological phases~\cite{Zhang2018,SM}.
More nontrivially, the $\mathbb{Z}_{2}$ phases can also be characterized by the highest-order BIS. For simplicity, we outline below the essential idea of proof for the first descendant $\mathbb{Z}_{2}$ topological phases. The full proof, including the $\mathbb{Z}^{(2)}_{2}$ invariants, can be found in Supplemental Material~\cite{SM}.

In general a $\mathbb{Z}_{2}$ topological phase in the AZ table can be derived as lower-dimensional descendant of the parent integer topological phase in the same symmetry class~\cite{Ryu2010,Qi2008} (see Tab.~\ref{tab:table1}). Thus the $d$D first descendant $\mathbb{Z}_{2}$ topological phase with $d'=d+1$ in Eq.~\eqref{eq:Hamiltonian} can be characterized by the parity of the integer invariant of a $(d+1)$D interpolation $H(\mathbf{k},\theta)=\sum^{d+1}_{i=0}h_{i}(\mathbf{k},\theta)\gamma_i$ of the same AZ class between $H(\mathbf{k},0)=H(\mathbf{k})$ and $H(\mathbf{k},\pi)=H_{\rm tr}(\mathbf{k})$ (a trivial reference Hamiltonian). Here $h_{i}(\mathbf{k},\theta)$ has the same parity as $h_{i}(\mathbf{k})$ to maintain the symmetries.

We shall characterize the $\mathbb{Z}_{2}$ phase based on the highest order BIS through the integer invariant $\mathcal{W}$ of $H(\mathbf{k},\theta)$. This is however not straightforward since we need to derive an index free of the parameter $\theta$ introduced in the interpolation.
We consider proper deformations for $H(\mathbf{k},\theta)$ such that one of its coefficients with odd parity, say $h'_{d+1}(\mathbf{k},\theta)$, vanishes at the plane $\theta=0$, while keeping the parity of integer invariant. Then the $n$-th order BIS $n\text{-BIS}=\{\mathbf{k},\theta\vert h'_{d+1}=h_{0}=\cdots=h_{n-2}=0\}$ becomes a $\theta$-independent $(d+1-n)$D symmetric subspace in the original BZ [Fig.~\ref{fig:figure1}(a)]. The interpolation $H(\mathbf{k},\theta)$ with multiple even coefficients can always be deformed into a trivial one~\cite{SM}, so a nontrivial $\mathbb{Z}_{2}$ topological phase has one coefficient being even, as considered here.

With the above process we can now express the $\mathbb{Z}_{2}$ index on the highest-order $(d+1)\text{-BISs}$ 
\begin{eqnarray}\label{eq:Z2_invariant}
    \nu^{(1)} &=& e^{\mathrm{i}\pi w^{(1)}},\\
    w^{(1)} &=& \frac{1}{2}\sum_{d\text{-BIS}_{j}}\sum^{N_j}_{i\in (d+1)\text{-BIS}}[\mathrm{sgn}(h_{\alpha_{i},i'})+\eta_{j}\mathrm{sgn}(h_{\alpha_{i},i''})],\nonumber
\end{eqnarray}
where the second summation is performed over the $N_{j}$ pairs of $0$D symmetric $(d+1)\text{-BIS}$ points $(i',i'')$ located on the $j$-th $1$D self-symmetric $d\text{-BIS}_{j}$ (or 1D $d\text{-BIS}_{j}$ pair) with $\eta_{j}=(-1)^{N_{j}}$ (or $-1$) [Fig.~\ref{fig:figure1}(b)]. The integer $w^{(1)}$, given by nonzero $h_{\alpha_{i}}$ ($\alpha_i={d},{d+1}$) of $H(\bold k)$ on the $(d+1)\text{-BIS}$ points, 
preserves the parity of the invariant $\mathcal{W}$ in the deformation.
For the second descendant $\mathbb{Z}_{2}$ topological phase with $d'=d+2$, by promoting it to a higher dimensional interpolation classified by $\mathbb{Z}^{(1)}_{2}$ and using the same techniques, we can obtain a similar invariant $\nu^{(2)}$, only with the modifications $n\text{-BIS}=\{\mathbf{k}\vert h_{i}=0,i=0,1,\dots,d'-3\}$ ($n=d'$ for the highest order BISs) and $h_{\alpha_{i}}\in\{h_{d},h_{d+1},h_{d+2}\}$~\cite{SM}.

With the Eq.~\eqref{eq:Z2_invariant} we have shown a profound result that the $d$D bulk topological phase~\eqref{eq:Hamiltonian} in all AZ symmetry classes can be characterized by a $0$D invariant defined on the highest-order BISs. Especially, the $\mathbb{Z}_{2}$ topology can be reduced to BIS points located at arbitrary discrete momenta of BZ, beyond the high symmetry points in the conventional characterization. This result updates the basic understanding of the tenfold classes of topological phases, and provides the foundation of the universal correspondence of equilibrium topological phases to quench dynamics, as studied below.

\emph{\bf Topological quench dynamics}.---We now study how to capture the dimension-reduced topology in the far-from-equilibrium quantum dynamics, induced by applying a constant magnetization along an arbitrary axis, say $\delta m_{0}\gamma_{0}$, to the Hamiltonian \eqref{eq:Hamiltonian} and quenching it from a large value ($\delta m_{0}\to\infty$) for the trivial initial state with density matrix $\rho_{0}$ to the topological regime with $\delta m_{0}=0$ at $t=0$; see Fig.~\ref{fig:figure1}(c). This quench dynamics can be quantified by the time-averaged spin polarization $\overline{\langle\boldsymbol{\gamma}(\mathbf{k})\rangle}_{0}\equiv\lim_{T\to\infty}\frac{1}{T}\int^{T}_{0}\mathrm{d}t\,\langle\boldsymbol{\gamma}(\mathbf{k},t)\rangle_{0}$, where $\langle\boldsymbol{\gamma}(\mathbf{k},t)\rangle_{0}=\mathrm{Tr}[\rho_{0}e^{\mathrm{i}Ht}\boldsymbol{\gamma}e^{-\mathrm{i}Ht}]$ evolves under the post-quench topological Hamiltonian $H(\mathbf{k})$.

We first identify the lowest-order BISs for post-quench Hamiltonian, i.e. the $(d-1)$D $(d'-d+1)\text{-BIS}_{(i)}$ with $h_{i}=0$, from quench dynamics [Fig.~\ref{fig:figure1}(d)]. The initial (pseudo)spin is perpendicular to $\mathbf{h}$-vector on the momenta where $h_{0}=0$, leading to a resonant spin-reversing dynamics and vanishing time-averaged spin polarization. Hence the $(d'-d+1)\text{-BIS}_{(0)}$ is formed by all the momenta $\mathbf{k}$ with $\overline{\langle\gamma_{\alpha}\rangle}_{0}=0$ for any $\alpha$. Further, on the momenta with $h_{i(\neq 0)}=0$ the spin polarization $\overline{\langle\gamma_{i}\rangle}_{0}$ vanishes, while all other components are in general nonzero, which identifies the $(d'-d+1)\text{-BIS}_{(i)}$. The parity of coefficients $h_{i}$ are determined from these BIS configurations.

\begin{figure}[b]
    \includegraphics{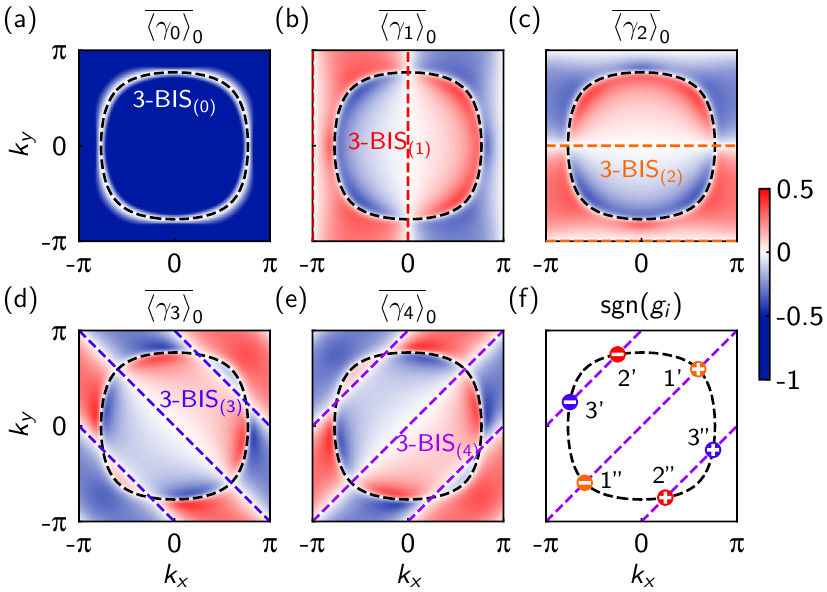}
    \caption{\label{fig:figure2}
    {\bf Characterizing $2$D topological phase in class AII.} (a)-(e) Time-averaged spin textures measured after a sudden quench from $m_{\mathrm{i}}=25t_{0}$ to $m_{\mathrm{f}}=-0.5t_{0}$. The $3\text{-BIS}_{(0)}$ is determined by $\overline{\langle\boldsymbol{\gamma}\rangle}_{0}=0$ (black dashed line). Additional lines with vanishing spin polarizations emerge in (b)-(e), giving the $3\text{-BIS}_{(i\neq 0)}$ (colored dashed lines), respectively. (f) The dynamical field $g_{i}$ on the $0$D $4\text{-BIS}$ constructed as the intersection points of $3\text{-BIS}_{(0)}$ and $3\text{-BIS}_{(4)}$. The red, orange and blue points represent the dynamical fields $g_{1}$, $g_{2}$ and $g_{3}$, respectively, with their signs indicated by $\pm$, characterizing the nontrivial $\mathbb{Z}_{2}$ topology. Here we set $t'_{0}=0.5t_{0}$ and $t_{\mathrm{so}}=0.2t_{0}$.
    }
\end{figure}

\begin{figure*}[t]
    \includegraphics{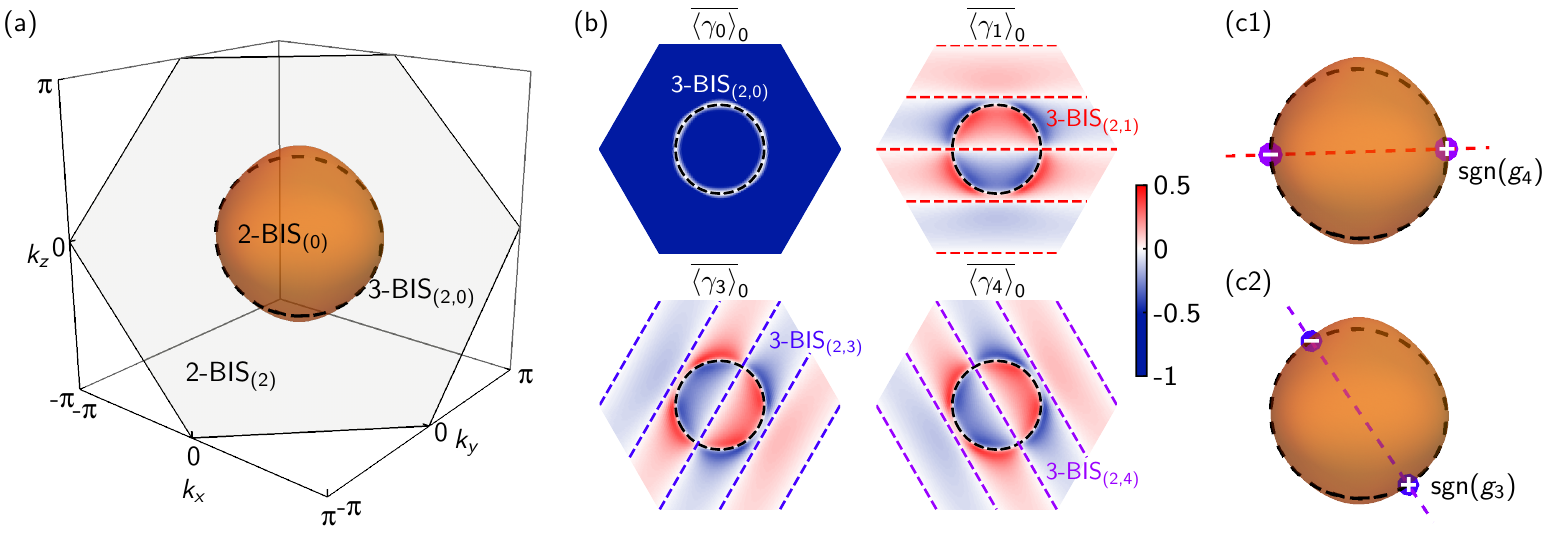}
    \caption{\label{fig:figure3}
    {\bf Dynamical detection of $3$D TR-invariant topological insulator in class AII.} (a) The $2\text{-BIS}_{(0)}$ (orange surface) and $2\text{-BIS}_{(2)}$ (gray plane) identified from the vanishing spin polarizations. Here we only show the $2\text{-BIS}_{(2)}$ having nonzero overlap with $2\text{-BIS}_{(0)}$. (b) Spin textures on the $2\text{-BIS}_{(2)}$ with vanishing $\overline{\langle\gamma_{2}\rangle}_{0}$. Here $3\text{-BIS}_{(2,i)}$ (colored dashed lines) denotes the intersection of $2\text{-BIS}_{(2)}$ and $2\text{-BIS}_{(i)}$. (c) The $0$D $4\text{-BIS}$ points constructed as the intersection of $3\text{-BIS}_{(2,0)}$ and $3\text{-BIS}_{(2,1)}$ (c1) or that of $3\text{-BIS}_{(2,0)}$ and $3\text{-BIS}_{(2,4)}$ (c2), on which the dynamical field $g_{i}$ is shown. These two different characterization schemes give the same dynamical invariant $\nu^{(1)}=-1$. Here we set $m_{\mathrm{i}}=25t_{0}$, $m_{\mathrm{f}}=1.8t_{0}$, and $t_{\mathrm{so}}=0.2t_{0}$.
    }
\end{figure*}

The higher-order $n$-BISs are determined dynamically as the intersections of the lowest-order ones $\bigcap_{i=0}^{n+d-d'-1}(d'-d+1)\text{-BIS}_{(i)}$. Of particular interest are the $1$D $(d'-1)\text{-BIS}$s and $0$D $d'\text{-BIS}$s. The topological index on the highest-order BISs is then captured through the dynamical field $g_{\alpha}(\mathbf{k})$, which quantifies the variation slope of $\overline{\langle\gamma_{\alpha}\rangle}_{0}$ across $d'\text{-BIS}$ at momentum $\mathbf{k}$, 
along the direction $k_{\perp}$ perpendicular to the $(d'-d+1)\text{-BIS}_{(0)}$ and pointing to the side with $h_{0}>0$~\cite{SM}. It follows from $\overline{\langle\gamma_{\alpha}\rangle}_{0}=-h_{\alpha}h_{0}/\mathbf{h}^2$ that $g_{\alpha}$ is proportional to $h_{\alpha}$. We then reach that $\mathcal{W}=\sum_{d\text{-BIS}_{j}}[\mathrm{sgn}(g_{d,R_{j}})-\mathrm{sgn}(g_{d,L_{j}})]/2$ for integer topological phases, and the first (second) descendant $\mathbb{Z}_{2}$ classifications with $d'=d+1$ (or $d+2$) can be characterized by the dynamical invariant
\begin{equation}\label{eq:dynamical_Z2_invariant}
    \nu^{(d'-d)} = \prod_{(d'-1)\text{-BIS}_{l}}\prod^{N_{l}}_{i\in d'\text{-BIS}}(-1)^{\frac{1}{2}[\mathrm{sgn}(g_{\alpha_{i},i'})+\eta_{l}\mathrm{sgn}(g_{\alpha_{i},i''})]},
\end{equation}
with nonzero $g_{\alpha_{i}}\in\{g_{d},\dots,g_{d'}\}$. 
The BIS configurations depend on which axis is chosen for quench, but the topological characterization is quench-axis independent.

The above results manifest a universal correspondence between the far-from-equilibrium quench dynamics and equilibrium topological phases for the complete AZ symmetry classes, as summarized in Tab.~\ref{tab:table1}. Moreover, this correspondence is not restricted on the deep trivial initial states, but also valid for generic quenches from an incompletely polarized phase~\cite{SM}.
For topological phases with the anti-unitary TR or chiral symmetries, our results are not affected by the so-called dynamical symmetry breaking~\cite{Gao2016,McGinley2018}, since the emergent dynamical topology on BISs is irrelevant to the instantaneous wavefunction.

\emph{\bf Dynamical detections}.---This theory provides the broadly applicable scheme with high feasibility to dynamically detect the complete AZ topological phases. Here we illustrate the $\mathbb{Z}_{2}$-classified examples, which, unlike the integer topological phases, have not been studied previously. We first consider a $1$D topological phase of class D with Hamiltonian $H(k)=(\mu-t_{0}\cos k)\sigma_{z}+\Delta\sin k\sigma_{x}+\Delta\sin 2k\sigma_{y}$, satisfying the PH symmetry $\sigma_{x}H^{*}(k)\sigma_{x}=-H(-k)$. Here $\mu$ and $\Delta$ resemble the chemical potential and pairing potential in superconductors, respectively, and $t_{0}$ is the hopping coefficient. This phase is characterized by the Chern-Simons invariant $\mathrm{CS}=\pm 1$ for $|\mu|\gtrless t_{0}$~\cite{Chiu2016}.
We quench the system from the trivial phase $\mu_{\mathrm{i}}\gg t_0$ to a topological regime with $|\mu_{\mathrm{f}}|<t_{0}$. Numerical results for $\mu_{\mathrm{f}}=0$ are shown in Fig.~\ref{fig:figure1}(e). The three time-averaged spin polarizations $\overline{\langle\sigma_{x,y,z}\rangle}_{z}$ are all nontrivial and differ from each other, indicating that the system has no chiral symmetry and no winding number can be defined. However, $\overline{\langle\boldsymbol{\sigma}\rangle}_{z}$ vanishes at two momenta $k_{L(R)}=\pm\pi/2$, manifesting the $0$D BIS points, on which the dynamical field $g_{x(y)}\sim\partial_{k_{\perp}}\overline{\langle\sigma_{x(y)}\rangle}_{z}$ is obtained. The opposite $g_{x}$ 
manifests a nontrivial $\mathbb{Z}_{2}$ invariant $\nu^{(1)}=-1$ for the postquench regime, while the component $g_{y}$ vanishes. For $\mu_{\mathrm{f}}\neq 0$, both $g_{x}$ and $g_{y}$ are nonzero, either of which can characterize the topological phase~\cite{SM}.

In the 2D case, we study a TR-invariant topological insulator in class AII, $H(\mathbf{k})=\mathbf{h}(\mathbf{k})\cdot\boldsymbol{\gamma}$, with $h_{0}(\mathbf{k})=m-t_{0}\sum_{i=x,y}\cos k_{i}-t'_{0}\sum_{i=1,2}\cos[k_{x}-(-1)^{i}k_{y}]$, $h_{i=1,2}(\mathbf{k})=t_{\mathrm{so}}\sin k_{i}$, and $h_{i=3,4}(\mathbf{k})=t_{\mathrm{so}}\sin[k_{x}-(-1)^{i}k_{y}]$. Here $t_{0},t'_{0}$ (or $t_{\mathrm{so}}$) denote the spin-conserved (-flipped) hopping coefficients, and $m$ is the effective magnetization. We take $\gamma_{0}=\mathbbm{1}\otimes\tau_{z}$, $\gamma_{1}=\sigma_{z}\otimes\tau_{x}$, $\gamma_{2}=\mathbbm{1}\otimes\tau_{y}$, $\gamma_{3}=\sigma_{x}\otimes\tau_{x}$ and $\gamma_{4}=\sigma_{y}\otimes\tau_{x}$, with $\sigma_{i}$ and $\tau_{i}$ being the Pauli matrices. The above Hamiltonian possesses the TR symmetry $-\mathrm{i}\sigma_{y}\otimes\mathbbm{1}\mathcal{K}$ ($\mathcal{K}$ is the complex conjugation operator) and is characterized by the Fu-Kane invariant $\mathrm{FK}=\mathrm{sgn}\{(m+2t'_{0})^{2}[(m-2t'_{0})^{2}-4t^{2}_{0}]\}$~\cite{Fu2006,Fu2007}.
Similar to the $1$D model, the quench process is performed by varying $m$ suddenly from $m_{\mathrm{i}}\gg 0$ to $|m_{\mathrm{f}}-2t'_{0}|<2t_{0}$ but $m_{\mathrm{f}}\neq -2t'_{0}$. From the numerical results in Figs.~\ref{fig:figure2}(a)-\ref{fig:figure2}(e) for $m_{\mathrm{f}}=-0.5t_{0}$ and $t'_{0}=0.5t_{0}$, the resonant spin-reversing dynamics with vanishing $\overline{\langle\boldsymbol{\gamma}\rangle}_{0}$ is obtained on the $3\text{-BIS}_{(0)}$, where 
the $4$D nature of the dynamical field $\mathbf{g}\propto-\partial_{k_{\perp}}\overline{\langle\boldsymbol{\gamma}\rangle}_{0}$ implies that the system has no integer topological numbers. Besides, the $3\text{-BIS}_{(i\neq 0)}$ can be read out from the corresponding spin texture $\overline{\langle\gamma_{i}\rangle}_{0}$, respectively. In Fig.~\ref{fig:figure2}(f), we use $3\text{-BIS}_{(0)}$ and $3\text{-BIS}_{(4)}$ to construct the $0$D $4\text{-BIS}$ beyond high symmetry points. Three pairs of symmetric points are obtained, to each of which we assign a nonzero dynamical field $g_{i}$,
giving the nontrivial $\mathbb{Z}_{2}$ invariant $\nu^{(2)}=-1$. In Supplemental Material~\cite{SM}, we also provide numerical results for quenching the $\gamma_{1}$ axis and a trivial example with BISs.

We finally also show application to the $3$D topological phase $H(\mathbf{k})=\mathbf{h}(\mathbf{k})\cdot\boldsymbol{\gamma}$ of class AII, with $h_{0}(\mathbf{k})=m-t_{0}\sum_{i=x,y,z}\cos k_{i}$, $h_{i}(\mathbf{k})=t_{\mathrm{so}}\sin(\sum_{j\neq i}k_{j}-k_{i})$ for $i=1,2,3$ and $h_{4}(\mathbf{k})=t_{\mathrm{so}}\sin(k_{x}+k_{y}+k_{z})$, which is characterized by Chern-Simons invariant~\cite{Chiu2016}. 
We consider the quench from initial trivial phase with $m_{\mathrm{i}}=25t_{0}$ to $m_{\mathrm{f}}=1.8t_{0}$, from which the $2\text{-BIS}_{(0)}$ and $2\text{-BIS}_{(2)}$ can be identified dynamically [Fig.~\ref{fig:figure3}(a)]. The spin textures on the $2\text{-BIS}_{(2)}$ are shown in Fig.~\ref{fig:figure3}(b), which gives the $3\text{-BIS}_{(2,i)}$, i.e. the intersection of $2\text{-BIS}_{(2)}$ and $2\text{-BIS}_{(i\neq2)}$. Here we provide two different but equivalent characterizations to illustrate the flexibility of our scheme. In Fig.~\ref{fig:figure3}(c1), the $0$D $4\text{-BIS}$ points are constructed as the intersections of $3\text{-BIS}_{(2,0)}$ and $3\text{-BIS}_{(2,1)}$, on which the dynamical field $g_{4}$ has opposite signs, manifesting the nontrivial dynamical $\mathbb{Z}_{2}$ invariant $\nu^{(1)}=-1$. Instead, in Fig.~\ref{fig:figure3}(c2) we use $3\text{-BIS}_{(2,0)}$ and $3\text{-BIS}_{(2,4)}$ to obtain the $4\text{-BIS}$ points. The corresponding dynamical field $g_{3}$ characterizes the same topology.

\emph{\bf Discussion}.--The uncovered universal correspondence shows insights into understanding both nonequilibrium quantum dynamics and equilibrium topological phases of the complete AZ tenfold classes. On one hand, this result provides a broadly applicable theory based on quantum dynamics for the characterization and detection of the complete AZ classes of topological phases; on the other hand, it shows that there are broad range of far-from-equilibrium quantum dynamics, usually being more complicated than equilibrium phases, which can be classified by topological theory. 
Finally, this study can be extended to topological phases classified by $230$ space groups~\cite{Slager2013} to build up an even broader theory to classify quantum dynamics and equilibrium topological phases, hence advances the research field widely.

This work was supported by National Natural Science Foundation of China (Grants No. 11825401, No. 11761161003, and No. 11921005), and the Strategic Priority Research Program of Chinese Academy of Science (Grant No. XDB28000000).



\renewcommand{\thesection}{S-\arabic{section}}
\setcounter{section}{0}  
\renewcommand{\theequation}{S\arabic{equation}}
\setcounter{equation}{0}  
\renewcommand{\thefigure}{S\arabic{figure}}
\setcounter{figure}{0}  
\renewcommand{\thetable}{S\Roman{table}}
\setcounter{table}{0}  

\onecolumngrid
\flushbottom


\begin{center}\large
\textbf{Supplementary Material}
\end{center}
\flushbottom

In this Supplemental Material, we first provide more details of the studies in the main text.

\section{Dirac Hamiltonian and the symmetry constraints}\label{sec:section1}

\subsection{Dirac Hamiltonian}

We consider a generic $d$D gapped topological phase in the Altland-Zirnbauer (AZ) symmetry classes~\cite{Chiu2016S} (Tab.~\ref{tab:AZ classes}) with the following Dirac Hamiltonian in terms of the Clifford algebra in irreducible representation
\begin{equation}\label{eq:Hamiltonian_S}
    H(\mathbf{k}) = \mathbf{h}(\mathbf{k})\cdot\boldsymbol{\gamma} = \sum_{i=0}^{d}h_{i}(\mathbf{k})\gamma_{i}+\sum_{i=d+1}^{d'}h_{i}(\mathbf{k})\gamma_{i},
\end{equation}
where $\mathbf{k}$ is the $d$D momentum. We have $d'=d$ for the integer topological phases and $d'=d+1$ (or $d+2$) for the first (second) descendant $\mathbb{Z}_{2}$ classifications; see Tab.~\ref{tab:AZ classes}. The $\gamma$ matrices satisfy $\gamma^{\dagger}_{i}=\gamma_{i}$ and the anti-commutation relation $\{\gamma_{i},\gamma_{j}\}=2\delta_{ij}$, mimicking a (pseudo) spin. These matrices can be constructed recursively as the tensor product of Pauli matrices. For example, we could define the matrices for $d'=2n$ as $\gamma^{(2n+1)}_{i}=\gamma^{(2n-1)}_{i}\otimes\sigma_{z}$ for $i=1,2,\dots,2n-2$, $\gamma^{(2n+1)}_{2n-1}=\mathbbm{1}_{2^{n-1}}\otimes\sigma_{x}$, $\gamma^{(2n+1)}_{2n}=\mathbbm{1}_{2^{n-1}}\otimes\sigma_y$, and $\gamma^{(2n+1)}_{0}=(\mathrm{i})^{n}\prod^{2n}_{i=1}\gamma^{(2n+1)}_{i}$. For $d'=2n-1$, we can identify $\gamma^{(2n)}_{i}$ with $\gamma^{(2n+1)}_{i}$ for $i=0,1,\dots,2n-1$ and leave out $\gamma^{(2n+1)}_{2n}$. The dimensionality of $\gamma$ matrices is given by $n_{d'}=2^{n}$, which is the minimal requirement for the $d$D gapped topological phases.

As examples, the $1$D first descendant $\mathbb{Z}_{2}$ and $2$D integer topological superconductors in class D have $\gamma_{0}=\sigma_z$, $\gamma_{1}=\sigma_{x}$ and $\gamma_{2}=\sigma_{y}$. On the other hand, the $1$D, $2$D and $3$D time-reversal-invariant topological superconductors in class DIII involves at least four bands and can be described by four Dirac matrices. The $2$D, $3$D and $4$D time-reversal-invariant topological insulators in class AII require five Dirac matrices.

\subsection{Symmetry constraints on the coefficients of Dirac Hamiltonian in real AZ classes}

The symmetries of Dirac Hamiltonian \eqref{eq:Hamiltonian_S} in real AZ classes are determined by the properties of $\gamma$ matrices and the coefficients $h_{i}(\mathbf{k})$. To facilitate the discussion, we start with the topological phases of $d'=2n$. For the corresponding $\gamma$ matrices, there exists a unitary matrix $C$ called the charge conjugation matrix~\cite{Das2014S}, satisfying $C^{-1}\gamma_{i}C=(-1)^{n}\gamma^{T}_{i}$ for $i=0,1,\dots,2n$ with $C^{T}=(-1)^{n(n+1)/2}C$. As an example, for the above recursive construction, 
the charge conjugation matrix is given by $C=\prod^{n}_{i=1}\gamma_{2i}$. This matrix is related to the anti-unitary symmetries. To see this, we define $U\equiv\gamma_{i_{1}}\gamma_{i_{2}}\cdots\gamma_{i_{\ell}}C$ with $i_{1}\neq i_{2}\neq\cdots\neq i_{\ell}$, satisfying $UU^{*}=(-1)^{n\ell}(-1)^{\ell(\ell-1)/2}(-1)^{n(n+1)/2}$. Acting on the $\gamma$ matrices, we have $U\gamma^{*}_{i}U^{-1}=(-1)^{n+\ell}\gamma_{i}$ for $i\notin\{i_{1},i_{2},\dots,i_{\ell}\}$, otherwise $U\gamma^{*}_{i}U^{-1}=(-1)^{n+\ell-1}\gamma_{i}$. Thus for proper coefficients $h_{i}$ and $\ell$, the operator $U\mathcal{K}$ represents the corresponding anti-unitary symmetry in the AZ ten-fold ways, i.e. the PH (TR) symmetry for odd (even) $n$; see Tab.~\ref{tab:AZ classes}. Here $\mathcal{K}$ is the complex conjugation operator.

We now consider the topological phase with $d'=2n-1$, which can be obtained from the one for $d'=2n$ by leaving out $\gamma_{2n}$. In addition to the symmetry $U\mathcal{K}$, the Hamiltonian \eqref{eq:Hamiltonian_S} now also possesses the chiral symmetry $\gamma_{2n}$ and another anti-unitary symmetry. This anti-unitary symmetry can be obtained by introducing the matrix $\tilde{U}\equiv U\gamma^{*}_{2n}$, which satisfies $\tilde{U}\gamma^{*}_{i}\tilde{U}^{-1}=-U\gamma^{*}_{i}U^{-1}$ and $\tilde{U}\tilde{U}^{*}=(-1)^{n+\ell}UU^{*}$ [or $\tilde{U}\tilde{U}^{*}=(-1)^{n+\ell-1}UU^{*}$] for odd $\ell$ and $2n\notin\{i_{1},i_{2},\dots,i_{\ell}\}$ [even $\ell$ and $2n\in\{i_{1},i_{2},\dots,i_{\ell}\}$]. From the AZ table, we know that the symmetry operator $\tilde{U}\mathcal{K}$ represents the TR (PH) symmetry for odd (even) $n$.

The important point is that according to the transformation of $\gamma$ matrices under symmetry operators $U\mathcal{K}$ and $\tilde{U}\mathcal{K}$, each coefficient $h_{i}$ of the Dirac Hamiltonian~\eqref{eq:Hamiltonian_S} in real AZ classes must be either odd or even with respect to $\mathbf{k}$. For example, if all coefficients $h_{i}$ are even, the Hamiltonian \eqref{eq:Hamiltonian_S} can describe a topological phase with the TR symmetry $U\mathcal{K}$ of $\ell=0$ for $d'=4n$. For $\ell=1$, if $h_{i_{1}}$ is an even function and all other coefficients $h_{i\neq i_{1}}$ are odd, then $U\mathcal{K}$ represents the PH (TR) symmetry for $d'=2n$ with odd (even) $n$, and we have $\tilde{U}\mathcal{K}$ being the additional TR (PH) symmetry for $d'=2n-1$ with odd (even) $n$. We would like to mention that the Dirac Hamiltonian \eqref{eq:Hamiltonian_S} also possesses the inversion symmetry $PH(\mathbf{k})P^{-1}=H(-\mathbf{k})$ with $P=\gamma_{i_{1}}\gamma_{i_{2}}\cdots\gamma_{i_{\ell}}$. However, the internal TR and/or PH symmetries still admit to define the $\mathbb{Z}$ or $\mathbb{Z}_{2}$ invariants.

\begin{table}[t]
    \setlength{\tabcolsep}{7pt}
    \begin{tabular}{ccccc|cccccccc}
        \hline
        \hline
        \multicolumn{5}{c|}{Symmetry} & \multicolumn{8}{c}{$d\mod8$}\tabularnewline

        $s$ & Class & $\mathcal{T}$ & $\mathcal{C}$ & $\mathcal{S}$ & $0$ & $1$ & $2$ & $3$ & $4$ & $5$ & $6$ & $7$\tabularnewline
        \hline
        $0$ & A & $0$ & $0$ & $0$ & $\mathbb{Z}$ & $0$ & $\mathbb{Z}$ & $0$ & $\mathbb{Z}$ & $0$ & $\mathbb{Z}$ & $0$\tabularnewline
        $1$ & AIII & $0$ & $0$ & $1$ & $0$ & $\mathbb{Z}$ & $0$ & $\mathbb{Z}$ & $0$ & $\mathbb{Z}$ & $0$ & $\mathbb{Z}$\tabularnewline
        \hline
        $0$ & AI & $+$ & $0$ & $0$ & $\mathbb{Z}$ & $0$ & $0$ & $0$ & $2\mathbb{Z}$ & $0$ & $\mathbb{Z}_{2}^{(2)}$ & $\mathbb{Z}_{2}^{(1)}$\tabularnewline
        $1$ & BDI & $+$ & $+$ & $1$ & $\mathbb{Z}_{2}^{(1)}$ & $\mathbb{Z}$ & $0$ & $0$ & $0$ & $2\mathbb{Z}$ & $0$ & $\mathbb{Z}_{2}^{(2)}$\tabularnewline
        $2$ & D & $0$ & $+$ & $0$ & $\mathbb{Z}_{2}^{(2)}$ & $\mathbb{Z}_{2}^{(1)}$ & $\mathbb{Z}$ & $0$ & $0$ & $0$ & $2\mathbb{Z}$ & $0$\tabularnewline
        $3$ & DIII & $-$ & $+$ & $1$ & $0$ & $\mathbb{Z}_{2}^{(2)}$ & $\mathbb{Z}_{2}^{(1)}$ & $\mathbb{Z}$ & $0$ & $0$ & $0$ & $2\mathbb{Z}$\tabularnewline
        $4$ & AII & $-$ & $0$ & $0$ & $2\mathbb{Z}$ & $0$ & $\mathbb{Z}_{2}^{(2)}$ & $\mathbb{Z}_{2}^{(1)}$ & $\mathbb{Z}$ & $0$ & $0$ & $0$\tabularnewline
        $5$ & CII & $-$ & $-$ & $1$ & $0$ & $2\mathbb{Z}$ & $0$ & $\mathbb{Z}_{2}^{(2)}$ & $\mathbb{Z}_{2}^{(1)}$ & $\mathbb{Z}$ & $0$ & $0$\tabularnewline
        $6$ & C & $0$ & $-$ & $0$ & $0$ & $0$ & $2\mathbb{Z}$ & $0$ & $\mathbb{Z}_{2}^{(2)}$ & $\mathbb{Z}_{2}^{(1)}$ & $\mathbb{Z}$ & $0$\tabularnewline
        $7$ & CI & $+$ & $-$ & $1$ & $0$ & $0$ & $0$ & $2\mathbb{Z}$ & $0$ & $\mathbb{Z}_{2}^{(2)}$ & $\mathbb{Z}_{2}^{(1)}$ & $\mathbb{Z}$\tabularnewline
        \hline
        \hline
    \end{tabular}
    \caption{\label{tab:AZ classes}
        Periodic table of topological phases in the AZ symmetry classes. The leftmost two columns denote the ten symmetry classes characterized by the presence or absence of time-reversal ($\mathcal{T}$), particle-hold ($\mathcal{C}$) and chiral ($\mathcal{S}$) symmetries of different types denoted by $\pm 1$. The entries $\mathbb{Z}$, $\mathbb{Z}_{2}$, $2\mathbb{Z}$ and $0$ represent the corresponding classification of the $d$D topological phases. Here $\mathbb{Z}^{(1,2)}_{2}$ denote the first and second descendant $\mathbb{Z}_{2}$ classifications.
        }
\end{table}

\section{Topological index defined on the highest order BISs}\label{sec:section2}

In this section, we introduce the dimensional reduction approach based on the higher order BISs and prove that an AZ class $d$D topological phase with an integer or $\mathbb{Z}_{2}$ invariant can be characterized by the $0$D invariant defined on the highest order BISs located at arbitrary discrete momenta of Brillouin zone (BZ).


\subsection{Review of dimensional reduction}

We first review the basics of the dimensional reduction~\cite{Qi2008S,Ryu2010S}, which states that all $\mathbb{Z}_{2}$ topological phases in the AZ symmetry classes can be derived as lower-dimensional descendants of parent $\mathbb{Z}$ topological phases in the same symmetry class, i.e. $\mathbb{Z}^{(2)}_{2}\leftarrow\mathbb{Z}^{(1)}_{2}\leftarrow\mathbb{Z}$; see Tab.~\ref{tab:AZ classes}.

\subsubsection{First descendant topological phases}

For the first descendant $\mathbb{Z}_{2}$ topological phase $H(\mathbf{k})$ with possible TR symmetry $\mathcal{T}$ and/or PH symmetry $\mathcal{C}$, we consider another trivial reference Hamiltonian $H_{\mathrm{tr}}(\mathbf{k})$ of the same symmetry class. Then a continuous interpolation $H(\mathbf{k},\theta)$ for $0\leq\theta\leq\pi$ between $H(\mathbf{k})$ and $H_{\mathrm{tr}}(\mathbf{k})$ can be constructed such that
\begin{equation}
    H(\mathbf{k},\theta=0)=H(\mathbf{k}),\qquad H(\mathbf{k},\theta=\pi)=H_{\mathrm{tr}}(\mathbf{k}).
\end{equation}
Since the topological space is simply connected, the continuous interpolation $H(\mathbf{k},\theta)$ is well defined. Together with its symmetry transformed partner, $H(\mathbf{k},\theta)=\mathcal{T}H(-\mathbf{k},-\theta)\mathcal{T}^{-1}$ and/or $H(\mathbf{k},\theta)=-\mathcal{C}H(-\mathbf{k},-\theta)\mathcal{C}^{-1}$ for $\theta\in[-\pi,0)$, the interpolation $H(\mathbf{k},\theta)$ obeys the same symmetries as $H(\mathbf{k})$ and is gapped for any $\theta\in[-\pi,\pi]$. Being in the same row of the tenfold way periodic table as $H(\mathbf{k})$ but with dimensionality increased by $1$, $H(\mathbf{k},\theta)$ now is characterized by an integer invariant denoted as $\mathcal{W}\in\mathbb{Z}$; see Tab.~\ref{tab:AZ classes}.

For any two distinct interpolations $H_{1}(\mathbf{k},\theta)$ and $H_{2}(\mathbf{k},\theta)$ between $H(\mathbf{k})$ and $H_{\mathrm{tr}}(\mathbf{k})$, the difference between the corresponding integer invariants is an even number~\cite{Qi2008S,Ryu2010S}. Therefore, the parity of the integer invariant $\mathcal{W}[H(\mathbf{k},\theta)]$ characterizes whether $H(\mathbf{k})$ and $H_{\mathrm{tr}}(\mathbf{k})$ belong to different first descendant $\mathbb{Z}_{2}$ topological phases and defines a $\mathbb{Z}_{2}$ invariant
\begin{equation}\label{eq:1st descendant Z2 invariant via dimensional reduction}
    \nu^{(1)}[H(\mathbf{k})]=e^{\mathrm{i}\pi \mathcal{W}[H(\mathbf{k},\theta)]}.
\end{equation}
The first descendant $\mathbb{Z}_{2}$ topological phase with $\nu^{(1)}[H(\mathbf{k})]=-1$ is topologically nontrivial and cannot be adiabatically deformed to $H_{\mathrm{tr}}(\mathbf{k})$ without gap close or breaking symmetries.

\subsubsection{Second descendant topological phases}

Using the same procedure, we can construct a higher-dimensional continuous interpolation $H(\mathbf{k},\phi)$ for $\phi\in[-\pi,\pi]$ between the second descendant $\mathbb{Z}_{2}$ topological phase $H(\mathbf{k})$ and the trivial one $H_{\mathrm{tr}}(\mathbf{k})$, such that $H(\mathbf{k},\phi=0)=H(\mathbf{k})$ and $H(\mathbf{k},\phi=\pi)=H_{\mathrm{tr}}(\mathbf{k})$. The interpolation $H(\mathbf{k},\phi)$ possesses the same symmetries as $H(\mathbf{k})$ and describes a first descendant $\mathbb{Z}_{2}$ topological phase in the same symmetry class.

For any two distinct interpolations $H_{1}(\mathbf{k},\phi)$ and $H_{2}(\mathbf{k},\phi)$, we further consider a interpolation between them
\begin{align}
    H(\mathbf{k},\phi,\theta=0) & =H_{1}(\mathbf{k},\phi),\quad H(\mathbf{k},\phi,\theta=\pi)=H_{2}(\mathbf{k},\phi),\nonumber \\
    H(\mathbf{k},\phi=0,\theta) & =H(\mathbf{k}),\qquad H(\mathbf{k},\phi=\pi,\theta)=H_{{\rm tr}}(\mathbf{k}).
\end{align}
We also require the interpolation $H(\mathbf{k},\phi,\theta)$ satisfying the same symmetries as $H(\mathbf{k})$, then $H(\mathbf{k},\phi,\theta)$ is characterized by an integer invariant. On the other hand, $H(\mathbf{k},\phi,\theta)$ can be viewed as the interpolation between the $\theta$-independent trivial first descendant $\mathbb{Z}_{2}$ phases $H(\mathbf{k},\phi=0,\theta)=H(\mathbf{k})$ and $H(\mathbf{k},\phi=\pi,\theta)=H_{\mathrm{tr}}(\mathbf{k})$ via the parameter $\phi$, leading to $e^{\mathrm{i}\pi \mathcal{W}[H(\mathbf{k},\phi,\theta)]}=+1$. Hence the interpolations $H_{1}(\mathbf{k},\phi)$ and $H_{2}(\mathbf{k},\phi)$ are topologically equivalent, and the interpolation $H(\mathbf{k},\phi)$ only depends on the end points $H(\mathbf{k})$ and $H_{\mathrm{tr}}(\mathbf{k})$.

For a trivial second descendant $\mathbb{Z}_{2}$ topological phase, we obviously have $\nu^{(1)}[H(\mathbf{k},\phi)]=+1$. The topological phase with $\nu^{(1)}[H(\mathbf{k},\phi)]=-1$ can be treated as $\mathbb{Z}_{2}$ nontrivial. Therefore, the second descendant $\mathbb{Z}_{2}$ topological phase $H(\mathbf{k})$ can be characterized by the invariant
\begin{equation}\label{eq:2nd descendant Z2 invariant via dimensional reduction}
    \nu^{(2)}[H(\mathbf{k})]=\nu^{(1)}[H(\mathbf{k},\phi)].
\end{equation}

\begin{figure}[b]
    \includegraphics{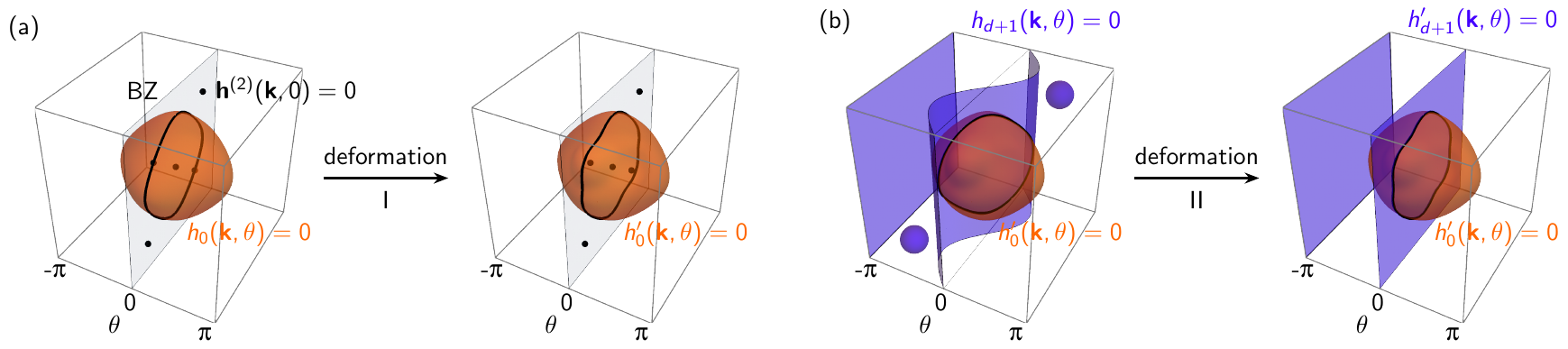}
    \caption{\label{fig:figureS1}
        Schematic illustration for the deformations of interpolation Hamiltonian $H(\mathbf{k},\theta)$. (a) The first deformation, in which the coefficient $h_{0}(\mathbf{k},\theta)$ is deformed into $h'_{0}(\mathbf{k},\theta)$ such that there is no momentum with $\mathbf{h}^{(2)}(\mathbf{k},0)\equiv(h_{1}(\mathbf{k},0),h_{2}(\mathbf{k},0),\dots,h_{d}(\mathbf{k},0))=0$ (black dots) on the $1\text{-BIS}'_{(0)}$ with $h'_{0}(\mathbf{k},\theta)=0$ (orange surface), i.e. $h^{\prime 2}_{0}(\mathbf{k},0)+\sum^{d}_{i=1}h^{2}_{i}(\mathbf{k},0)>0$ for all momenta $\mathbf{k}$. (b) The second deformation $h_{d+1}\to h'_{d+1}$. These $1\text{-BIS}_{(d+1)}$'s with $h_{d+1}(\mathbf{k},\theta)=0$ and not intersected with the plane $\theta=0$ (blue spheres) come in pairs and can be deformed to vanish. The open (blue) surface crossing the point $(\mathbf{k},\theta)=0$ is deformed to the plane $\theta=0$.
        }
\end{figure}

\subsection{Integer topological phases}

We now proceed to show the dimensional reduction approach based on the higher order BISs and start with the integer topological phases. The BIS is a key concept in the so-called bulk-surface duality~\cite{Zhang2018S,Yu2020S}, showing that characterizing a $d$D bulk integer topological phase can reduce to a lower-dimensional invariant defined on the higher order BISs. Here we briefly review the main results. For the more details, we refer the readers to Refs.~\cite{Zhang2018S,Yu2020S}.

For integer topological phases~\eqref{eq:Hamiltonian_S} with $d'=d$, we first decompose the Hamiltonian vector field $\mathbf{h}$ into two parts, $h_{0}$ and $\mathbf{h}^{(1)}\equiv(h_{1},h_{2},\dots,h_{d})$. Physically, $h_{0}$ characterizes the decoupled energy dispersions, of which the band crosses give the $(d-1)$D first order BISs, $1\text{-BIS}\equiv\{\mathbf{k}\in\mathrm{BZ}\vert h_{0}(\mathbf{k})=0\}$. On the other hand, the term $\mathbf{h}^{(1)}$ describes the coupling between these decoupled bands and opens a topological gap on the $1\text{-BIS}$s. In Ref.~\cite{Zhang2018S}, we show that the $d$D bulk integer topology can reduce to the nontrivial topology of vector field $\mathbf{h}^{(1)}$ on the $1\text{-BIS}$s, captured by the following $(d-1)$D winding number
\begin{equation}
    \mathcal{W} = \frac{\Gamma(d/2)}{2\pi^{d/2}(d-1)!}\int_{1\text{-BIS}}\hat{\mathbf{h}}^{(1)}[\mathrm{d}\hat{\mathbf{h}}^{(1)}]^{d-1},
\end{equation}
where $\Gamma(x)$ is the Gamma function and the hat represents the normalization. For further reduction, this winding number can be treated as the integer invariant of a $(d-1)$D gapped Hamiltonian defined on the $1\text{-BIS}$s, namely $\tilde{H}(\mathbf{k})=\mathbf{h}^{(1)}(\mathbf{k})\cdot\tilde{\boldsymbol{\gamma}}$, where $\tilde{\gamma}_{i}$ is the corresponding $\gamma$ matrix in the $(d-1)$D space. We can introduce the second-order BISs as $2\text{-BIS}\equiv\{\mathbf{k}\in 1\text{-BIS}\vert h_{1}(\mathbf{k})=0\}$, then the $(d-1)$D integer topology of $\tilde{H}$ can be reduced to the $(d-2)$D winding number of vector field $\mathbf{h}^{(2)}\equiv(h_{2},h_{3},\dots,h_{d})$ on the $2\text{-BIS}$s according to the bulk-surface duality.

Repeating this procedure, we can further introduce the $n$-th order BISs as $n\text{-BIS}\equiv\{\mathbf{k}\in\mathrm{BZ}\vert h_{0}(\mathbf{k})=h_{1}(\mathbf{k})=\cdots=h_{n-1}(\mathbf{k})=0\}$. Correspondingly, the integer invariant $\mathcal{W}$ now reduces to the winding number of the $(d+1-n)$D vector field $\mathbf{h}^{(n)}\equiv(h_{n},h_{n+1},\dots,h_{d})$ on the $(d-n)$D $n\text{-BIS}$s~\cite{Yu2020S}:
\begin{equation}\label{eq:integer invariant on nth order BIS_S}
    \mathcal{W} = \frac{\Gamma[(d+1-n)/2]}{2\pi^{(d+1-n)/2}(d-n)!}\int_{n\text{-BIS}}\hat{\mathbf{h}}^{(n)}[\mathrm{d}\hat{\mathbf{h}}^{(n)}]^{d-n}.
\end{equation}
This formula shows a correspondence between the bulk integer topological phase and the characterization on the higher order BISs. Particularly, on the $0$D highest order $d\text{-BIS}$s, we have
\begin{equation}\label{eq:integer invariant on the highest order BIS_S}
    \mathcal{W} = \frac{1}{2}\sum_{d\text{-BIS}_{j}}[\mathrm{sgn}(h_{d,R_{j}})-\mathrm{sgn}(h_{d,L_{j}})],
\end{equation}
where $d\text{-BIS}_{j}$ comprises the left (right) boundary $d\text{-BIS}$ point $L_{j}$ ($R_{j}$) of the $j$-th segment with $h_{d-1}<0$ on the $1$D $(d-1)\text{-BIS}$s. Note that the construction of the $n\text{-BIS}$s and the corresponding vector field $\mathbf{h}^{(n)}$ is actually not unique, which can be understood as a relabeling for the Hamiltonian coefficients, i.e. $(h_{0},h_{1},\dots,h_{d'})\to(h_{i_{0}},h_{i_{1}},\dots,h_{i_{d'}})$.

\subsection{The first descendant $\mathbb{Z}_{2}$ topological phases}

We now turn to the first descendant $\mathbb{Z}_{2}$ topological phase \eqref{eq:Hamiltonian_S} with $d'=d+1$ and show how to derive a $\mathbb{Z}_{2}$ topological index defined on the highest order BISs. According to the dimensional reduction, the invariant $\nu^{(1)}$ of the $d$D Dirac Hamiltonian \eqref{eq:Hamiltonian_S} can be determined by the parity of the integer invariant of the following $(d+1)$D interpolation
\begin{equation}
    H(\mathbf{k},\theta)=\mathbf{h}(\mathbf{k},\theta)\cdot\boldsymbol{\gamma}=\sum^{d+1}_{i=0}h_{i}(\mathbf{k},\theta)\gamma_{i},
\end{equation}
with $H(\mathbf{k},\theta=0)=H(\mathbf{k})$ and $H(\mathbf{k},\theta=\pi)$ being trivial (e.g. being fully polarized). Here the parities of coefficients $h_{i}(\mathbf{k},\theta)$ with respect to $(\mathbf{k},\theta)$ are required to be the same as those for $h_{i}(\mathbf{k})$ to maintain the symmetries of $H(\mathbf{k})$.

Naively, the $\mathbb{Z}_{2}$ invariant on the $0$D highest order BISs can be directly obtained by the $\mathcal{W}$ invariant of $H(\mathbf{k},\theta)$; see Eq.~\eqref{eq:integer invariant on the highest order BIS_S}. However, the above scheme involves the explicit form of interpolation and depends on an additional parameter $\theta$. To derive an index free of the interpolation, we can further make appropriate deformations for $H(\mathbf{k},\theta)$ without changing the parity of its integer invariant, such that the corresponding higher order BISs become the $\theta$-independent subspaces in the original BZ.

\subsubsection{First deformation}

As the first step, we choose an arbitrary component, say $h_{0}(\mathbf{k},\theta)$, and deform the interpolation $H(\mathbf{k},\theta)$ into
\begin{equation}\label{eq:Z21_deformation1}
    H'_{1}(\mathbf{k},\theta)=h'_{0}(\mathbf{k},\theta)\gamma_{0}+\sum^{d+1}_{i=1}h_{i}(\mathbf{k},\theta)\gamma_{i},
\end{equation}
where $h'_{0}(\mathbf{k},\theta)$ has the same parity as $h_{0}(\mathbf{k},\theta)$ and satisfies $h^{\prime 2}_{0}(\mathbf{k},\theta=0)+\sum^{d}_{i=1}h^{2}_{i}(\mathbf{k},\theta=0)>0$ for all momenta $\mathbf{k}$; see Fig.~\ref{fig:figureS1}(a). For this, there must be at least one even function $h_{i}(\mathbf{k},\theta)$ for $0\leq i\leq d$. The deformed interpolation $H'_{1}(\mathbf{k},\theta)$ belongs to the same symmetry class as $H(\mathbf{k},\theta)$ but may have different integer invariant. We can show that compared with $H(\mathbf{k},\theta)$, the integer invariant of $H'_{1}(\mathbf{k},\theta)$ changes at most by an even number, which does not affect the characterization of the first descendant $\mathbb{Z}_{2}$ topological phases, namely we still have $\nu^{(1)}[H(\mathbf{k})]=e^{\mathrm{i}\pi \mathcal{W}[H'_{1}(\mathbf{k},\theta)]}$. The proof is given as follows.

According to the bulk-surface duality~\cite{Zhang2018S}, the integer invariant of $H'_{1}(\mathbf{k},\theta)$ can also be interpreted as the total topological charges enclosed by the first-order BISs $1\text{-BIS}'_{(0)}\equiv\{(\mathbf{k},\theta)\vert h'_{0}(\mathbf{k},\theta)=0\}$, namely
\begin{equation}
    \mathcal{W}[H'_{1}(\mathbf{k},\theta)]=\sum_{(\mathbf{k}_{*},\theta_{*})\in\mathcal{V}_{1\text{-BIS}'_{(0)}}}Q_{(\mathbf{k}_{*},\theta_{*})},
\end{equation}
where $\mathcal{V}_{1\text{-BIS}'_{(0)}}$ denotes the region enclosed by $1\text{-BIS}'_{(0)}$ with $h'_{0}<0$, and the topological charge $Q_{(\mathbf{k}_{*},\theta_{*})}$ is defined as the monopole charge of the vector field $\mathbf{h}^{(1)}= (h_{1},h_{2},\dots,h_{d+1})$ at the node point $(\mathbf{k}_{*},\theta_{*})$, i.e. $\mathbf{h}^{(1)}(\mathbf{k}_{*},\theta_{*})=0$. Since each coefficient $h_{i}$ is either even or odd, the topological charges are symmetrically distributed. That is, if we have a topological charge at $(\mathbf{k}_{*},\theta_{*})$ with value $Q_{(\mathbf{k}_{*},\theta_{*})}$, then there is also a charge at $(-\mathbf{k}_{*},-\theta_{*})$, satisfying $|Q_{(-\mathbf{k}_{*},-\theta_{*})}|=|Q_{(\mathbf{k}_{*},\theta_{*})}|$. For the same reason, the first-order BISs $1\text{-BIS}'_{(0)}$ also remain symmetric with respect to $(\mathbf{k},\theta)$ under the deformation $h_{0}(\mathbf{k},\theta)\to h'_{0}(\mathbf{k},\theta)$. We see that if the topological charge at $(\mathbf{k}_{*},\theta_{*})$ passes through the $1\text{-BIS}'_{(0)}$ during the deformation, so does the charge at $(-\mathbf{k}_{*},-\theta_{*})$, which at most changes the integer invariant of $H(\mathbf{k},\theta)$ by an even number. This proves the above statement. Note that the deformed $1\text{-BIS}'_{(0)}$ shall not cross the high symmetry points, at which the value of the topological charge may be odd.

\subsubsection{Second deformation}

In the next step, we further make the deformation $h_{d+1}(\mathbf{k},\theta)\to h'_{d+1}(\mathbf{k},\theta)$ such that the Hamiltonian \eqref{eq:Z21_deformation1} becomes
\begin{equation}\label{eq:Z21_deformation2}
    H'_{2}(\mathbf{k},\theta)=h'_{0}(\mathbf{k},\theta)\gamma_{0}+\sum^{d}_{i=1}h_{i}(\mathbf{k},\theta)\gamma_{i}+h'_{d+1}(\mathbf{k},\theta)\gamma_{d+1},
\end{equation}
where $h'_{d+1}(\mathbf{k},\theta)$ has the same parity as $h_{d+1}(\mathbf{k},\theta)$. This deformation is used to eliminate the $\theta$-dependence of the invariant $\nu^{(1)}$. We explain this in terms of topological charges $Q'$ defined by the vector field $\mathbf{h}^{(1)\prime}\equiv (h'_{0},h_{1},h_{2},\dots,h_{d})$ and the first-order BISs $1\text{-BIS}'_{(d+1)}\equiv\{(\mathbf{k},\theta)\vert h'_{d+1}(\mathbf{k},\theta)=0\}$. Similar to the first deformation, the topological charges $Q'$ are also symmetrically distributed, and the $1\text{-BIS}'_{(d+1)}$'s remain symmetric with respect to $(\mathbf{k},\theta)$. For this, we only need to concern the $1\text{-BIS}_{(d+1)}$ intersected with the plane $\theta=0$. Other BISs must come in pairs [see Fig.~\ref{fig:figureS1}(b)] and can be deformed to vanish without changing the parity of integer invariant $\mathcal{W}[H'_{1}(\mathbf{k},\theta)]$ by using the similar arguments of the first deformation.

Note that from the deformation \eqref{eq:Z21_deformation1}, there is no topological charge $Q'$ with $\mathbf{h}^{(1)\prime}=0$ in the plane $\theta=0$. If the coefficient $h_{d+1}$ is an even function of $(\mathbf{k},\theta)$, the $1\text{-BIS}_{(d+1)}$'s are closed surfaces in the space $(\mathbf{k},\theta)$ and can be deformed to be close to the plane $\theta=0$ via the deformation $h_{d+1}\to h'_{d+1}$, such that there is no topological charge $Q'$ enclosed by the $1\text{-BIS}'_{(d+1)}$'s, manifesting a trivial integer invariant $\mathcal{W}[H'_{2}(\mathbf{k},\theta)]$. As there is at least one coefficient $h_{i}$ for $0\leq i\leq d$ being even, this suggests that {\it the first descendant $\mathbb{Z}_{2}$ topological phase $H(\mathbf{k})$ with multiple even coefficients is always trivial}. The only possible nontrivial phases are these with only one even coefficient. In this work, we focus on the latter case. For the odd function $h_{d+1}(\mathbf{k},\theta)$, the open surface $1\text{-BIS}_{(d+1)}$ crossing the point $(\mathbf{k},\theta)=0$ can be deformed into the plane $\theta=0$ without affecting the parity of $\mathcal{W}[H'_{1}(\mathbf{k},\theta)]$, namely we have
\begin{equation}
    1\text{-BIS}'\equiv 1\text{-BIS}'_{(d+1)}=\mathrm{BZ};
\end{equation}
see Fig.~\ref{fig:figureS1}(b). Note that for the fully polarized Hamiltonian $H(\mathbf{k},\theta=\pi)$, the $1\text{-BIS}_{(d+1)}$ plane at $\theta=\pi$ can be ignored as it does not contribute to the integer invariant of $H'_{2}(\mathbf{k},\theta)$. Now the topological invariant for the first descendant $\mathbb{Z}_{2}$ topological phases can be expressed as
\begin{equation}
    \nu^{(1)}[H(\mathbf{k})]=e^{\mathrm{i}\pi \mathcal{W}[H'_{2}(\mathbf{k},\theta)]}.
\end{equation}

\begin{figure}[ht]
    \includegraphics{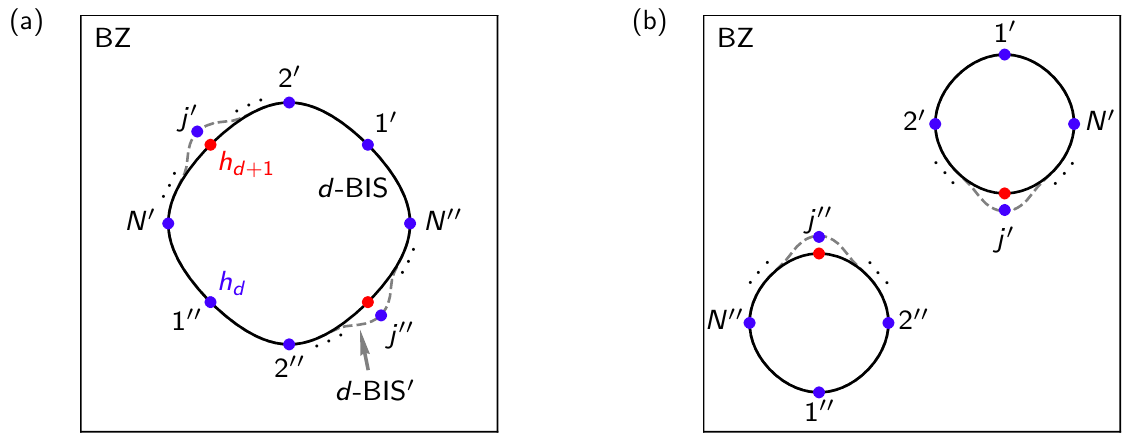}
    \caption{\label{fig:figureS2}
        The symmetric $(d+1)$-th order BIS point pairs $(i',i'')$ on the self-symmetric $1$D $d$-th order BIS (a) and the symmetric $d$-th order BIS pair (b). The black lines represent the original $d\text{-BIS}$, on which the $(d+1)\text{-BIS}$ points are denoted by the dots. Near the $j$-th $(d+1)\text{-BIS}$ point pair with vanishing $h_{d}$ but nonzero $h_{d+1}$ (red dot), the $d\text{-BIS}$ is deformed into the $d\text{-BIS}'$ (gray dashed line) via deformation $h_{0}\to h'_{0}$, such that the $j$-th $(d+1)\text{-BIS}'$ point pair has nonzero $h_{d}$ (blue dot).
        }
\end{figure}

\subsubsection{Topological index $\nu^{(1)}$ defined on the highest order BISs}

With the above deformations, the higher order BISs $n\text{-BIS}'\equiv\{(\mathbf{k},\theta)\vert h'_{d+1}(\mathbf{k},\theta)=h'_{0}(\mathbf{k},\theta)=h_{1}(\mathbf{k},\theta)=\cdots=h_{n-2}(\mathbf{k},\theta)=0\}=\{\mathbf{k}\in\mathrm{BZ}\vert h'_{0}(\mathbf{k})=h_{1}(\mathbf{k})=\cdots=h_{n-2}(\mathbf{k})=0\}$ now become the subspaces of the original BZ, on which the integer invariant $\mathcal{W}[H'_{2}(\mathbf{k},\theta)]$ in terms of vector field $\mathbf{h}^{(n)}\equiv(h_{n-1},h_{n},\dots,h_{d})$ is $\theta$-independent. However, this invariant requires the deformation $h_{0}\to h'_{0}$ to guarantee that the coefficients $h_{i}$ for $i=1,2,\dots,d$ cannot be zero simultaneously on the higher order BISs. To derive a formula for the topological index $\nu^{(1)}$ without deformation, we rewrite Eq.~\eqref{eq:integer invariant on the highest order BIS_S} into the following form
\begin{equation}\label{eq:Z21_0D_deformation_pair}
    \mathcal{W}[H'_{2}(\mathbf{k},\theta)]=\sum_{d\text{-BIS}'_{j}}\sum^{N_{j}}_{i\in(d+1)\text{-BIS}'}\frac{(-1)^{i}}{2}[\mathrm{sgn}(h_{d,i'})+\eta_{j}\mathrm{sgn}(h_{d,i''})],
\end{equation}
where $d\text{-BIS}'_{j}$ denotes the $j$-th $1$D $d\text{-BIS}'$, and the second summation is performed over the $N_{j}$ $0$D symmetric $(d+1)\text{-BIS}'$ point pairs on the $d\text{-BIS}'_{j}$, ordered by $(i',i'')$; see Fig.~\ref{fig:figureS2}. Here $\eta_{j}$ is given by $(-1)^{N_{j}}$ for the self-symmetric $d\text{-BIS}'_{j}$ [Fig.~\ref{fig:figureS2}(a)] or takes the value $-1$ for the symmetric $d\text{-BIS}'$ pair [Fig.~\ref{fig:figureS2}(b)].

We notice that the deformed $(d+1)\text{-BIS}'$ points are obtained from the original $(d+1)\text{-BIS}=\{\mathbf{k}\vert h_{0}(\mathbf{k})=h_{1}(\mathbf{k})=\cdots=h_{d-1}(\mathbf{k})\}$ via a slight deformation $h_{0}\to h'_{0}$, which in general does not affect the corresponding order $i'$, $i''$. Since $h_{d}$ and $h_{d+1}$ cannot be zero simultaneously on certain pair $(i',i'')$ of the $(d+1)\text{-BIS}$ points, we can make the following replacement for Eq.~\eqref{eq:Z21_0D_deformation_pair}
\begin{equation}
    \frac{1}{2}[\mathrm{sgn}(h_{d,i'})\pm\mathrm{sgn}(h_{d,i''})]_{(d+1)\text{-BIS}'}\to\frac{1}{2}[\mathrm{sgn}(h_{\alpha_{i},i'})\pm\mathrm{sgn}(h_{\alpha_{i},i''})]_{(d+1)\text{-BIS}}
\end{equation}
if $h_{d}$ and $h_{d+1}$ have the same parity; see Fig.~\ref{fig:figureS2}. Here $h_{\alpha_{i}}\in\{h_{d},h_{d+1}\}$ is nonzero on the original $(d+1)\text{-BIS}$ points labeled by $i'$ and $i''$. Correspondingly, we have
\begin{equation}
    \mathcal{W}[H'_{2}(\mathbf{k},\theta)]\to w^{(1)}=\frac{1}{2}\sum_{d\text{-BIS}_{j}}\sum^{N_{j}}_{i\in(d+1)\text{-BIS}}[\mathrm{sgn}(h_{\alpha_{i},i'})+\eta_{j}\mathrm{sgn}(h_{\alpha_{i},i''})],
\end{equation}
where the factor $(-1)^{i}$ has been ignored without affecting the parity of $w^{(1)}$. Under the above replacements, each term in Eq.~\eqref{eq:Z21_0D_deformation_pair} changes only by $0,\pm 2$, manifesting that $w^{(1)}$ has the same parity as $\mathcal{W}[H'_{2}(\mathbf{k},\theta)]$, which does not affect the characterization of the first descendant $\mathbb{Z}_{2}$ topological phases.

Finally, we arrive at the $0$D topological index $\nu^{(1)}$ defined on the highest order BISs for the Hamiltonian \eqref{eq:Hamiltonian_S} without knowing the explicit interpolation and requiring the deformation:
\begin{equation}
    \nu^{(1)} = e^{\mathrm{i}\pi w^{(1)}} = \prod_{d\text{-BIS}_{j}}\prod^{N_{j}}_{i\in(d+1)\text{-BIS}}(-1)^{\frac{1}{2}[\mathrm{sgn}(h_{\alpha_{i},i'})+\eta_{j}\mathrm{sgn}(h_{\alpha_{i},i''})]},
\end{equation}
where $h_{\alpha_{i}}\in\{h_{d},h_{d+1}\}$ is nonzero on the corresponding $(d+1)\text{-BIS}$ point pair $(i',i'')$, with $h_{d}$ and $h_{d+1}$ having the same parity. Note that for each of the symmetric $d\text{-BIS}$ pair, the integer invariant \eqref{eq:integer invariant on nth order BIS_S} has the same absolute value [Fig.~\ref{fig:figureS2}(b)], which has a trivial contribution to the index $\nu^{(1)}$. The only possible nontrivial contribution comes from the $(d+1)\text{-BIS}$ points on the self-symmetric BISs [Fig.~\ref{fig:figureS2}(a)].

\subsection{The second descendant $\mathbb{Z}_{2}$ topological phases}

For the second descendant $\mathbb{Z}_{2}$ topological phase~\eqref{eq:Hamiltonian_S} with $d'=d+2$, the topological index $\nu^{(2)}$ can be derived similarly. Consider the Dirac Hamiltonian $H(\mathbf{k})=\sum_{i=0}^{d+2}h_{i}(\mathbf{k})\gamma_{i}$, we first construct a $(d+1)$D interpolation
\begin{equation}
    H(\mathbf{k},\phi)=\mathbf{h}(\mathbf{k},\phi)\cdot\boldsymbol{\gamma}=\sum_{i=0}^{d+2}h_{i}(\mathbf{k},\phi)\gamma_{i},
\end{equation}
with $H(\mathbf{k},\phi=0)=H(\mathbf{k})$ and $H(\mathbf{k},\phi=\pi)$ being fully polarized. Here the parities of the coefficients $h_{i}(\mathbf{k},\phi)$ with respect to $(\mathbf{k},\phi)$ are the same as those for $h_{i}(\mathbf{k})$ with respect to $\mathbf{k}$. The interpolation $H(\mathbf{k},\phi)$ belongs to the same symmetry class as $H(\mathbf{k})$. According to the dimensional reduction, we have $\nu^{(2)}[H(\mathbf{k})]=\nu^{(1)}[H(\mathbf{k},\phi)]$. As argued above, the possible nontrivial Hamiltonian $H(\mathbf{k},\phi)$ has only one even coefficient. Without loss of generality, we assume $h_{d},h_{d+1}$ and $h_{d+2}$ to be odd.

From the results of the first descendant $\mathbb{Z}_{2}$ topological phases, the topological invariant $\nu^{(1)}[H(\mathbf{k},\phi)]$ can be determined by the parity of the winding number of vector field $(h'_{0}(\mathbf{k},\phi),h_{1}(\mathbf{k},\phi),\dots,h_{d+1}(\mathbf{k},\phi))$ on the $1\text{-BIS}'=\mathrm{BZ}\times(-\pi,\pi]$, where $h_{0}(\mathbf{k},\phi)$ has been deformed to guarantee $h^{\prime 2}_{0}(\mathbf{k},\phi)+\sum_{i=1}^{d+1}h^2_{i}(\mathbf{k},\phi)>0$ for all $(\mathbf{k},\phi)$. This winding number can be treated as the integer invariant of a $(d+1)$D gapped Hamiltonian defined on the $\mathrm{BZ}\times(-\pi,\pi]$, namely $\tilde{H}(\mathbf{k},\phi)=h'_{0}(\mathbf{k},\phi)\tilde{\gamma}_{0}+\sum_{i=1}^{d+1}h_{i}(\mathbf{k},\phi)\tilde{\gamma}_{i}$, with $\tilde{\gamma}_{i}$ being the corresponding $\gamma$ matrices in the $(d+1)$D space. To eliminate the interpolation parameter $\phi$, we further utilize the first and second deformations of the first  descendant $\mathbb{Z}_{2}$ topological phases to deform $\tilde{H}(\mathbf{k},\phi)$ into the following form
\begin{equation}
    \tilde{H}'(\mathbf{k},\phi)=h''_{0}(\mathbf{k},\phi)\tilde{\gamma}_{0}+\sum^{d}_{i=1}h_{i}(\mathbf{k},\phi)\tilde{\gamma}_{i}+h'_{d+1}(\mathbf{k},\phi)\tilde{\gamma}_{d+1}
\end{equation}
such that $h^{\prime\prime 2}_{0}(\mathbf{k},\phi=0)+\sum^{d}_{i=1}h^{2}_{i}(\mathbf{k},\phi=0)>0$ for all $\mathbf{k}\in\mathrm{BZ}$ and the open surface with $h'_{d+1}(\mathbf{k},\phi)=0$ is given by the plane $\phi=0$. This deformation keeps the parity of the integer invariant of  $\tilde{H}(\mathbf{k},\phi)$ unchanged, hence we have $\nu^{(2)}[H(\mathbf{k})]=e^{\mathrm{i}\pi \mathcal{W}[\tilde{H}'(\mathbf{k},\phi)]}$.

Using the same techniques for the first descendant $\mathbb{Z}_{2}$ topological phases, we find that the topological index $\nu^{(2)}$ on the highest order BISs can be expressed as
\begin{equation}
    \nu^{(2)} = \prod_{(d+1)\text{-BIS}_{j}}\prod^{N_{j}}_{i\in(d+2)\text{-BIS}}(-1)^{\frac{1}{2}[\mathrm{sgn}(h_{\alpha_{i},i'})+\eta_{j}\mathrm{sgn}(h_{\alpha_{i},i''})]},
\end{equation}
where we have $n\text{-BIS}\equiv\{\mathbf{k}\in\mathrm{BZ}\vert h_{0}(\mathbf{k})=h_{1}(\mathbf{k})=\cdots=h_{n-3}(\mathbf{k})=0\}$, $n\geq 3$, and $h_{\alpha_{i}}\in\{h_{d},h_{d+1},h_{d+2}\}$ are nonzero on the corresponding $(d+2)\text{-BIS}$ point pair $(i',i'')$, with $h_{d},h_{d+1}$ and $h_{d+2}$ having the same parity. Here $\eta_{j}$ is given by $(-1)^{N_{j}}$ with $N_{j}$ being the total number of pairs $(i',i'')$ for the self-symmetric $(d+1)\text{-BIS}_{j}$, and is $-1$ for the symmetric $(d+1)\text{-BIS}_{j}$ pair; see Fig.~\ref{fig:figureS2}. Interestingly, the topological index $\nu^{(2)}$ has the same form as the index $\nu^{(1)}$ when reduced to the highest order BISs, providing a unified method to characterize the $\mathbb{Z}_{2}$ topological phases.

\begin{figure}[b]
    \includegraphics{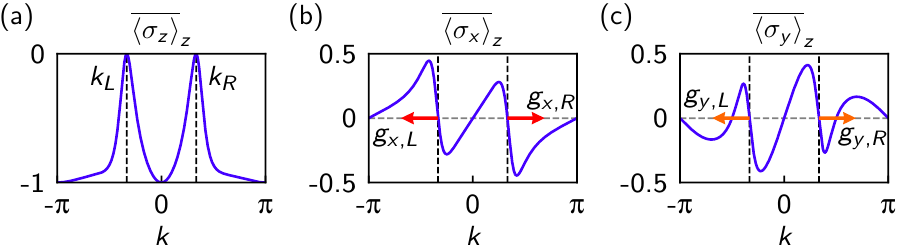}
    \caption{\label{fig:figureS3}
        Numerical results for $1$D topological superconductor of class D. The time-averaged spin polarizations are obtained by quenching the system from $\mu_{\mathrm{i}}=25t_{0}$ to $\mu_{\mathrm{f}}=0.5t_{0}$. The vanishing polarization $\overline{\langle\boldsymbol{\sigma}\rangle}_{z}$ gives the BIS points $k_{L,R}$ (black dashed lines). The opposite dynamical field $g_{x}$ (red arrows) and $g_{y}$ (orange arrows) on the BIS points are shown in (b) and (c) respectively, manifesting a nontrivial dynamical invariant $\nu^{(1)}=-1$. Here we set $\Delta=0.2t_{0}$.
    }
\end{figure}

\section{Numerical Results}\label{sec:section3}

In this section, we provide more numerical results, which serve as a supplement to the main text.

\subsection{1D class D topological phase with both nonzero $g_{x}$ and $g_{y}$}

We first consider the $1$D topological phase of class D studied in the main text, $H(k)=(\mu-t_{0}\cos k)\sigma_{z}+\Delta\sin k\sigma_{x}+\Delta\sin 2k\sigma_{y}$, satisfying the PH symmetry $\sigma_{x}H^{*}(k)\sigma_{x}=-H(-k)$. Here $t_{0}$ is the hopping coefficient, and the parameters $\mu$ and $\Delta$ may denote the chemical potential and pairing order, respectively, if considering superconductors. 
This phase is characterized by the Chern-Simons invariant $\mathrm{CS}=\exp(-\int^{\pi}_{-\pi}\mathrm{d}k\,\langle u_{-}\vert\partial_{k}u_{-}\rangle)$ with $\vert u_{-}\rangle$ being the ground state~\cite{Chiu2016S}, which equals $\pm 1$ for $|\mu|\gtrless t_{0}$. We quench the system from a deep trivial state $\rho_{0}$ with $\mu_{\mathrm{i}}=25t_{0}$ to a topologically nontrivial regime with $\mu_{\mathrm{f}}=0.5t_{0}$. The numerical results are shown in Fig.~\ref{fig:figureS3}. We can see that the time-averaged spin polarization $\overline{\langle\boldsymbol{\sigma}\rangle}_{z}\equiv\lim_{T\to\infty}\frac{1}{T}\int^{T}_{0}\mathrm{d}t\,\mathrm{Tr}[\rho_{0}e^{\mathrm{i}Ht}\boldsymbol{\sigma}e^{-\mathrm{i}Ht}]$ vanishes at two BIS points $k_{L(R)}=\pm\pi/3$. The corresponding dynamics field components $g_{x/y}=-(1/\mathcal{N}_{k})\partial_{k_{\perp}}\overline{\langle\sigma_{x/y}\rangle}_{z}$ are both nonzero and point to opposite directions on the left and right BIS points. Here $\mathcal{N}_{k}$ is a normalization factor, and $k_{\perp}$ is perpendicular to the BIS and points to the side with $h_{z}>0$. This $\mathbb{Z}_{2}$ topological phase can be characterized by any one of these two nonzero dynamical field components, giving the same dynamical invariant $\nu^{(1)}=-1$.

\begin{figure}[b]
    \includegraphics{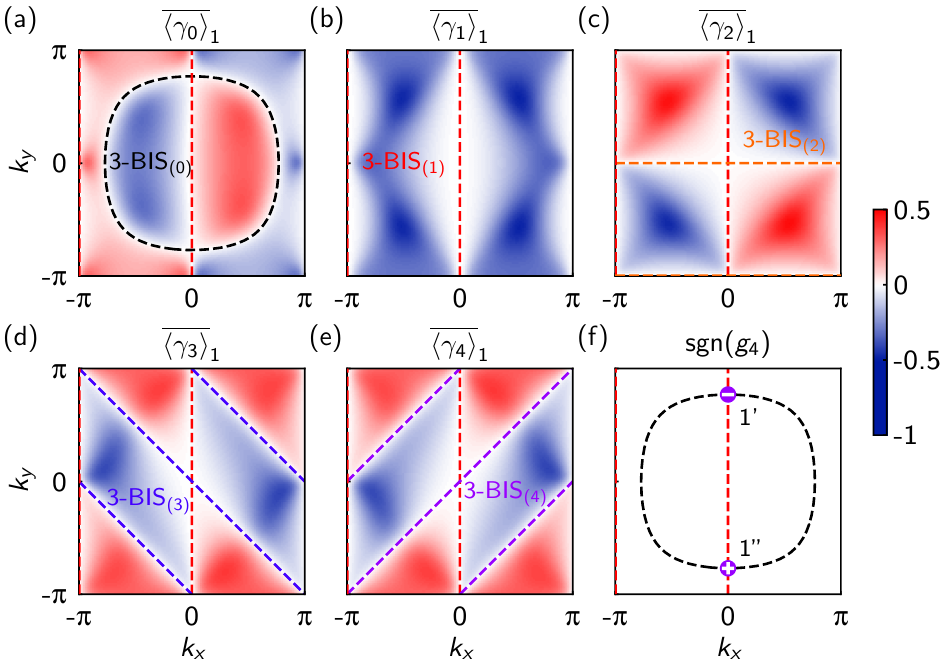}
    \caption{\label{fig:figureS4}
        Characterizing $2$D TR-invariant topological insulator in class AII by quenching the $\gamma_{1}$ axis. (a)-(e) Time-averaged spin textures for the quantum dynamics induced by quenching $\delta m_{1}$ from $100t_{0}$ to $0$. The $3\text{-BIS}_{(1)}$ is determined by $\overline{\langle\boldsymbol{\gamma}\rangle}_{1}=0$ (red dashed line). Besides, additional lines with vanishing $\overline{\langle\gamma_{i}\rangle}_{1}$ emerge in the spin textures (a) and (c)-(e), giving the $3\text{-BIS}_{(i\neq 1)}$ (colored dashed lines). (f) The dynamical field $g_{4}$ on the $0$D $4\text{-BIS}$ constructed as the intersection of $3\text{-BIS}_{(0)}$ and $3\text{-BIS}_{(1)}$. The opposite signs indicated by $\pm$ characterize the nontrivial $\mathbb{Z}_{2}$ topology with $\nu^{(2)}=-1$. Here we set $m=-0.5t_{0}$, $t'_{0}=0.5t_{0}$ and $t_{\rm so}=0.2t_{0}$.
    }
\end{figure}

\subsection{Characterizing the 2D class AII TR-invariant topological insulator by quenching the $\gamma_{1}$ axis}\label{sec:section3B}

Next, we consider the 2D class AII TR-invariant topological insulator, $H(\mathbf{k})=[m-t_{0}\cos k_{x}-t_{0}\cos k_{y}-t'_{0}\cos(k_{x}+k_{y})-t'_{0}\cos(k_{x}-k_{y})]\gamma_{0}+t_{\rm so}\sin k_{x}\gamma_{1}+t_{\rm so}\sin k_{y}\gamma_{2}+t_{\rm so}\sin(k_{x}+k_{y})\gamma_{3}+t_{\rm so}\sin(k_{x}-k_{y})\gamma_{4}$. Here $t_{0},t'_{0}$ (or $t_{\mathrm{so}}$) denote the spin-conserved (-flipped) hopping coefficients, and $m$ is the effective magnetization. The $\gamma$ matrices are given as $\gamma_{0}=\mathbbm{1}\otimes\tau_{z}$, $\gamma_{1}=\sigma_{z}\otimes\tau_{x}$, $\gamma_{2}=\mathbbm{1}\otimes\tau_{y}$, $\gamma_{3}=\sigma_{x}\otimes\tau_{x}$ and $\gamma_{4}=\sigma_{y}\otimes\tau_{x}$, with $\sigma_{i}$ and $\tau_{i}$ being the Pauli matrices. This Hamiltonian possesses the TR symmetry $\mathcal{T}=-\mathrm{i}\sigma_{y}\otimes\mathbbm{1}\mathcal{K}$ ($\mathcal{K}$ is the complex conjugation operator) and is characterized by the Fu-Kane invariant $\mathrm{FK}=\prod_{\mathbf{K}}\mathrm{Pf}[B(\mathbf{K})]/\sqrt{\det[B(\mathbf{K})]}=\mathrm{sgn}\{(m+2t'_{0})^{2}[(m-2t'_{0})^{2}-4t^{2}_{0}]\}$, where $\mathbf{K}$ runs over high symmetry points and the sewing matrix reads $B^{\alpha\beta}=\langle u_{\alpha}(-\mathbf{k})\vert\mathcal{T}u_{\beta}(\mathbf{k})\rangle$ for occupied states $\vert u_{\alpha,\beta}\rangle$~\cite{Fu2006S,Fu2007S}.

In the main text, we have shown the results for quenching the $\gamma_{0}$ axis. Here we further consider the quench dynamics along the $\gamma_{1}$ axis, whose coefficient is an odd function. The quench process is performed by adding a constant term $\delta m_{1}\gamma_{1}$ to the Hamiltonian $H(\mathbf{k})$, with $\delta m_{1}\to\infty$ for the pre-quench trivial state and $\delta m_{1}=0$ for the post-quench topological phase. Note that this quench process is between different symmetry classes as the term $\delta m_{1}\gamma_{1}$ breaks the symmetries of $H(\mathbf{k})$. The numerical results are shown in Figs.~\ref{fig:figureS4}(a)-\ref{fig:figureS4}(e) for $m=-0.5t_{0}$, from which the $3\text{-BIS}_{(1)}$ with $h_{1}=0$ is characterized by the momenta with vanishing $\overline{\langle\boldsymbol{\gamma}\rangle}_{1}$. Besides, the $3\text{-BIS}_{(i)}$ for $i=0,2,3,4$ are given by the additional lines with vanishing values in the corresponding spin texture $\overline{\langle\gamma_{i}\rangle}_{1}$. In Fig.~\ref{fig:figureS4}(f), we show the $0$D $4\text{-BIS}$ points constructed from the $3\text{-BIS}_{(0)}$ and $3\text{-BIS}_{(1)}$, on which the opposite dynamical field $g_{4}\sim-\partial_{k_{\perp}}\overline{\langle\gamma_{4}\rangle}_{1}$ with $k_{\perp}$ perpendicular to the $3\text{-BIS}_{(1)}$ gives the nontrivial dynamical invariant $\nu^{(2)}=-1$. This result is consistent with the quench dynamics along the $\gamma_{0}$ axis, manifesting the flexibility of our dynamical characterization scheme.

\subsection{2D trivial phase with BISs}

\begin{figure}[t]
    \includegraphics{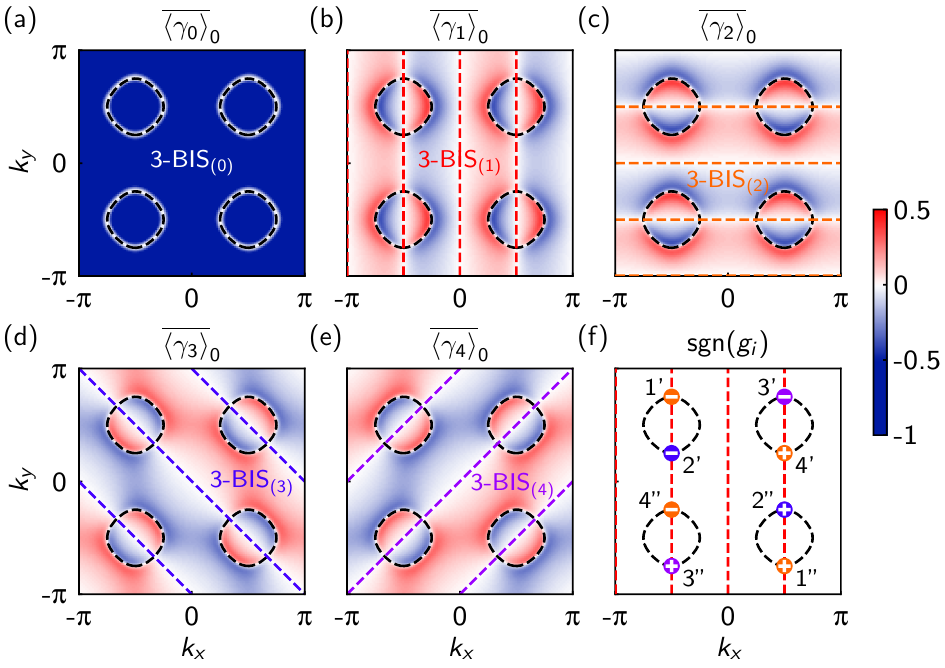}
    \caption{\label{fig:figureS5}
        $2$D trivial phase with BISs. (a)-(e) Time-averaged spin textures show the
        third order BISs $3\text{-BIS}_{(i)}$ for $i=0,1,\dots,4$. (f) The dynamical field $g_{i}$ on the $0$D $4\text{-BIS}$ points obtained from $3\text{-BIS}_{(0)}$ and $3\text{-BIS}_{(1)}$. The orange, blue and purple points represent the dynamical field $g_{2}$, $g_{3}$, $g_{4}$ respectively, with its signs indicated by $\pm$, which lead to a trivial dynamical invariant $\nu^{(2)}=+1$. Here the quench dynamics is induced by varying $m_{\rm i}=25t_{0}$ to $m_{\rm f}=-t_{0}$, with $t_{\rm so}=0.2t_{0}$.
    }
\end{figure}

We now consider a $2$D trivial example in class AII with the following Hamiltonian, $H(\mathbf{k})=\mathbf{h}\cdot\boldsymbol{\gamma}=(m-t_{0}\cos 2k_{x}-t_{0}\cos 2k_{y})\gamma_{0}+t_{\rm so}\sin 2k_{x}\gamma_{1}+t_{\rm so}\sin 2k_{y}\gamma_{2}+t_{\rm so}\sin(k_{x}+k_{y})\gamma_{3}+t_{\rm so}\sin(k_{x}-k_{y})\gamma_{4}$. The $\gamma$ matrices are the same as those in the above subsection. The corresponding Fu-Kane invariant is always trivial. After suddenly quenching $m$ from $m_{\rm i}=25t_{0}$ to $m_{\rm f}=-t_{0}$, the time-averaged spin texture $\overline{\langle\boldsymbol{\gamma}\rangle}_{0}$ exhibits two symmetric $3\text{-BIS}_{(0)}$ pairs with $h_{0}=0$ [see Figs.~\ref{fig:figureS5}(a)-\ref{fig:figureS5}(e)]. Meanwhile, in the spin texture $\overline{\langle\gamma_{i\neq 0}\rangle}_{0}$, we can also read out the corresponding $3\text{-BIS}_{(i)}$ with $h_{i}=0$. The dynamical fields $g_{i}\propto-\partial_{k_{\perp}}\overline{\langle\gamma_{i}\rangle}_{0}$ on the $4\text{-BIS}$ constructed as the intersection of $3\text{-BIS}_{(0)}$ and $3\text{-BIS}_{(1)}$ are shown in Fig.~\ref{fig:figureS5}(f), from which we obtain the dynamical invariant $\nu^{(2)}=\prod^{4}_{i=1}(-1)^{\frac{1}{2}[\mathrm{sgn}(g_{\alpha_{i},i'})-\mathrm{sgn}(g_{\alpha_{i},i''})]}=+1$, manifesting the trivial $\mathbb{Z}_{2}$ phase. Here $k_{\perp}$ is perpendicular to the $3\text{-BIS}_{(0)}$ and points to the side with $h_{0}>0$.

\section{Universal dynamical correspondence in the shallow quenches}\label{sec:section4}

In this section, we generalize the universal correspondence between the far-from-equilibrium quench dynamics and equilibrium 
topological phases in the AZ table established in the main text to the shallow quench regime, where the initial trivial state is not fully polarized.

We consider the quench process realized by tuning an arbitrary axis $\gamma_{0}$. For this, we write $\tilde{h}_{0}(\mathbf{k})=\delta m_{0}+h_{0}(\mathbf{k})$, where $\delta m_{0}$ is an additional constant magnetization. The shallow quench is triggered by suddenly changing $\delta m_{0}$ from a finite value to zero, with an incompletely polarized initial state $\rho_{0}(\mathbf{k})=(1/n_{d'})[I-\hat{H}_{\mathrm{pre}}(\mathbf{k})]$, where $\hat{H}_{\mathrm{pre}}(\mathbf{k})$ is the flattened pre-quench Hamiltonian $H(\mathbf{k})+\delta m_{0}\gamma_{0}$. After quench, the quantum dynamics is governed by the unitary evolution under the post-quench topological phase $H(\mathbf{k})$, giving the spin polarization $\langle\boldsymbol{\gamma}(t)\rangle_{0}=\mathrm{Tr}[\rho_{0}e^{\mathrm{i}Ht}\boldsymbol{\gamma}e^{-\mathrm{i}Ht}]$. As the initial state now is momentum-dependent, the resonant spin-reversing transition $\langle\boldsymbol{\gamma}\rangle\to-\langle\boldsymbol{\gamma}\rangle$ does not occur on the $(d'-d+1)\text{-BIS}_{(0)}$ with $h_{0}=0$ in general (see the deep quench case in main text). Instead, the time-averaged spin polarizations
\begin{equation}
    \overline{\langle\gamma_{i}(\mathbf{k})\rangle}_{0}=\lim_{T\to\infty}\frac{1}{T}\int^{T}_{0}\mathrm{d}t\,\langle\gamma_{i}(\mathbf{k},t)\rangle_{0}=h_{i}(\mathbf{k})\mathrm{Tr}[\rho_{0}(\mathbf{k})H(\mathbf{k})]/\mathbf{h}^{2}(\mathbf{k})
\end{equation}
manifest a $(d-1)$D momentum subspace with $\mathrm{Tr}[\rho_{0}H]=0$, where the resonant transition happens and we have $\overline{\langle\boldsymbol{\gamma}(\mathbf{k})\rangle}_{0}=0$. We refer to this hypersurface as the dynamical band-inversion surface (dBIS):
\begin{equation}
    \text{dBIS}\equiv\{\mathbf{k}\in\mathrm{BZ}\vert\overline{\langle\gamma_{i}(\mathbf{k})\rangle}_{0}=0,\forall i\}.
\end{equation}
In the deep quench limit $\delta m_{0}\to\infty$, the dBIS coincides with the $(d'-d+1)\text{-BIS}_{(0)}$. Now in the shallow quench, the $(d'-d+1)\text{-BIS}_{(i)}$ with $h_{i}=0$ for $i=0,1,\dots,d'$ can be obtained as
\begin{equation}
    (d'-d+1)\text{-BIS}_{(i)}=\{\mathbf{k}\in\mathrm{BZ}\vert\overline{\langle\gamma_{i}(\mathbf{k})\rangle}_{0}=0\}-\mathrm{dBIS}.
\end{equation}
The parity of coefficients $h_{i}$ can be determined from the configurations of these BISs. If $(d'-d+1)\text{-BIS}_{(i)}$ contains the momentum $\mathbf{k}=0$, then the coefficient $h_{i}$ is odd with respect to $\mathbf{k}$, otherwise it is even. For the $\mathbb{Z}_{2}$ topological phases, once there exist multiple even coefficients, one can immediately conclude that the Hamiltonian $H(\mathbf{k})$ is trivial. For the possible nontrivial $\mathbb{Z}_{2}$ phases, we assume $h_{i\geq d}$ to be odd without loss of generality. 

\begin{figure}[t]
    \includegraphics{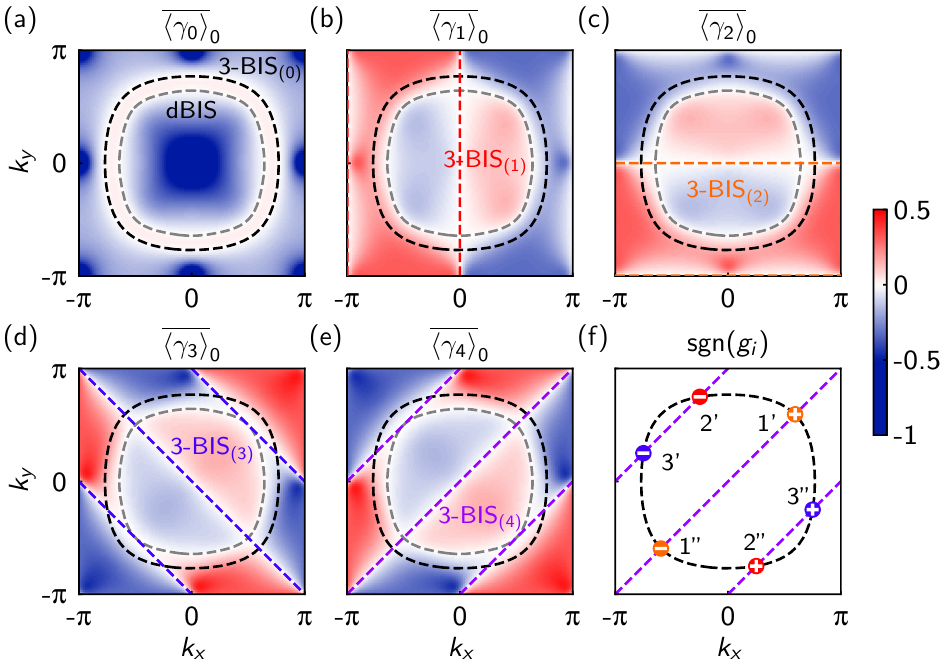}
    \caption{\label{fig:figureS6}
        Dynamical detection of $2$D class AII TR-invariant topological insulator via a shallow quench. (a)-(e) Time-averaged spin textures $\overline{\langle\gamma_{i}\rangle}_{0}$. Here we take $t_{\rm so}=t_{0}$ and tune $m$ from $4t_{0}$ to $-0.5t_{0}$, with $t'_{0}=0.5t_{0}$. A ring-shaped structure (dashed gray line) with vanishing polarization emerges in all the spin textures, which characterizes the dBIS. In each spin texture $\overline{\langle\gamma_{i}\rangle}_{0}$, there are additional lines with vanishing values, representing the corresponding $3\text{-BIS}_{(i)}$ with $h_{i}=0$. Here we also plot $3\text{-BIS}_{(0)}$ in $\overline{\langle\gamma_{i>0}\rangle}_{0}$ for convenience. (f) Generalized dynamical field $g_{i}$ on the $4\text{-BIS}=3\text{-BIS}_{(0)}\cap 3\text{-BIS}_{(4)}$. The red, orange and blue points represent the dynamical field $g_{1}$, $g_{2}$, $g_{3}$ respectively. The configuration of the dynamical field remains the same as the one in deep quench (see main text) and characterizes the nontrivial topology.
    }
\end{figure}

To capture the dimension-reduced topology on the highest order BISs, the $1$D $(d'-1)\text{-BIS}$ and $0$D $d'\text{-BIS}$ can be constructed in the same way as in the deep quench, namely
\begin{equation}
    (d'-1)\text{-BIS}=\bigcap^{d-2}_{i=0}(d'-d+1)\text{-BIS}_{(i)}\qquad\mathrm{and}\qquad d'\text{-BIS}=\bigcap^{d-1}_{i=0}(d'-d+1)\text{-BIS}_{(i)}.
\end{equation}
In contrast to the deep quench, the spin polarization $\overline{\langle\boldsymbol{\gamma}\rangle}_{0}$ in shallow quench may not always vanish on the $d'\text{-BIS}$ points, except for the case where the $d'\text{-BIS}$ point also belongs to the dBIS. For this, we generalize the dynamical field $g_{i}(\mathbf{k}\in d'\text{-BIS})$ as
\begin{equation}\label{eq:generalized_dynamical_field}
    g_{i}(\mathbf{k})\equiv\begin{cases}
        -(1/\mathcal{N}_{\mathbf{k}})\partial_{k_{\perp}}\overline{\langle\gamma_{i}(\mathbf{k})\rangle}_{0} & \text{if $\mathbf{k}$ is also on dBIS}, \\
        \zeta_{\mathbf{k}}\overline{\langle\gamma_{i}(\mathbf{k})\rangle}_{0}/\mathcal{N}_{\mathbf{k}} & \text{otherwise},
    \end{cases}
\end{equation}
for $i=d,d+1,\dots,d'$. Here $k_{\perp}$ is perpendicular to the dBIS (i.e., the contour $\mathrm{Tr}[\rho_{0}H]$) and points to the side with negative $\zeta_{\mathbf{k}}$, and $\zeta_{\mathbf{k}}$ denotes the sign of $\mathrm{Tr}[\rho_{0}(\mathbf{k})H(\mathbf{k})]$. For two adjacent regions separated by the dBIS, $\zeta_{\mathbf{k}}$ will have different signs in general. Note that only the relative signs are relevant for the $\mathbb{Z}_{2}$ topological phases. One can easily check that the dynamical field $g_{i}$ is proportional to $h_{i}$ on the $d'\text{-BIS}$. Then we have $\mathcal{W}=\sum_{d\text{-BIS}_{j}}[\mathrm{sgn}(g_{d,R_{j}})-\mathrm{sgn}(g_{d,L_{j}})]/2$ for the integer topological phases with $d'=d$, and the first (second) descendant $\mathbb{Z}_{2}$ topological phases with $d'=d+1$ (or $d+2$) still can be characterized by the following dynamical invariant
\begin{equation}\label{eq:dynamical_Z2_invariant_S}
    \nu^{(d'-d)} = \prod_{(d'-1)\text{-BIS}_{l}}\prod^{N_{l}}_{i\in d'\text{-BIS}}(-1)^{\frac{1}{2}[\mathrm{sgn}(g_{\alpha_{i},i'})+\eta_{l}\mathrm{sgn}(g_{\alpha_{i},i''})]}
\end{equation}
with $g_{\alpha_{i}}\in\{g_{d},g_{d+1},\dots,g_{d'}\}$ being nonzero on the corresponding $d'\text{-BIS}$ point pair $(i',i'')$. 

As an example, we consider the $2$D TR-invariant topological insulator in Sec.~\ref{sec:section3B}. Here we set $t_{\mathrm{so}}=t_{0}$, and the shallow quench is performed by tuning $m$ from $m_{\rm i}=4t_{0}$ to $m_{\rm f}=-0.5t_{0}$. The time-averaged spin textures are shown in Figs.~\ref{fig:figureS6}(a)-\ref{fig:figureS6}(e), from which both the dBIS and $3\text{-BIS}_{(i)}$ for $i=0,1,\dots,4$ can be identified. In Fig.~\ref{fig:figureS6}(f), we calculate the generalized dynamical field $g_{i}$ [see Eq.~\eqref{eq:generalized_dynamical_field}] on the $0$D $4\text{-BIS}$ constructed as the intersection points of $3\text{-BIS}_{(0)}$ and $3\text{-BIS}_{(4)}$. It is seen that the corresponding configuration of dynamical field remains the same as the one in the deep quench (see main text), which gives the dynamical invariant $\nu^{(2)}=-1$ and characterizes the nontrivial $\mathbb{Z}_{2}$ topological phase.


\end{document}